\documentclass[iop,twocolappendix]{emulateapj}

\usepackage{graphicx}
\usepackage{amssymb}
\usepackage{amsmath}
\usepackage{natbib}
\usepackage{longtable}
\usepackage{footnote}
\usepackage{helvet}
\usepackage{color}
\usepackage{afterpage}
\usepackage{float}
\usepackage{soul}
\usepackage[toc,page]{appendix}
\usepackage{chngcntr}
\usepackage{placeins}

\usepackage[breaklinks, colorlinks, citecolor=blue]{hyperref} 

\slugcomment{Accepted to The Astrophysical Journal January 25, 2016}

\begin{document}

\def\edit#1{#1}

\newcommand{\HCOrot}{HCO$^+$ (3--2)}
\newcommand{\HCO}{HCO$^+$}
\newcommand{\NtDrot}{N$_2$D$^+$ (3--2)}
\newcommand{\NtD}{N$_2$D$^+$}
\newcommand{\NtHrot}{N$_2$H$^+$ (1--0)}
\newcommand{\NtH}{N$_2$H$^+$}
\newcommand{\ammonia}{NH$_3$}
\newcommand{\deltaV}{${\delta}_v$}
\newcommand{\deltaVstar}{${\delta}_v^*$}
\newcommand{\deltaT}{${\delta}_T$}
\newcommand{\VLSR}{$V_{\mbox{\tiny{LSR}}}$}
\newcommand{\Vthin}{$V_\mathrm{thin}$}
\newcommand{\Vthick}{$V_\mathrm{thick}$}
\newcommand{\Vcent}{$V_\mathrm{thin,\mathrm{HCO^+}}$}
\newcommand{\sigmathin}{${\sigma}_\mathrm{thin}$}
\newcommand{\Vin}{$V_\mathrm{in}$}
\newcommand{\Vmod}{$V_\mathrm{mod}$}
\newcommand{\Msun}{M$_{\odot}$}
\def\plus{\texttt{+}}
\newcommand{\Atemp}{$T_A^*$}
\newcommand{\TMB}{$T_\mathrm{MB}$}
\newcommand{\Tex}{$T_\mathrm{ex}$}
\newcommand{\kms}{$\mathrm{km\,s^{-1}}$}
\newcommand{\redchisq}{$\bar{{\chi}}^2$}
\newcommand{\microm}{${\mu}m${\,}}
\newcommand{\rhoPCC}{${\rho}_{\mbox{\tiny{PCC}}}$}
\newcommand{\Hill}{HILL5}
\newcommand{\MJ}{$M_\mathrm{J}$}
\newcommand{\MMJ}{$M/M_\mathrm{J}$}
\newcommand{\tk}{$T_\mathrm{K}$}
\newcommand{\tatm}{\tau_\mathrm{atm}} 
\def\peter#1{{\bf \textcolor{red}{[#1]}}}
\def\comments#1{\noindent{\bf \textcolor{green}{[#1]}}} 

\makeatletter
\setlength{\@fptop}{0pt}
\makeatother

\newcommand\T{\rule{0pt}{2.6ex}} 
\newcommand\B{\rule[-1.2ex]{0pt}{0pt}} 

\title{Contraction signatures toward dense cores in the Perseus molecular cloud}
\shorttitle{Contraction Signatures of Perseus Dense Cores}

\author{J. L. Campbell\altaffilmark{1,2}}
\author{R. K. Friesen\altaffilmark{2}}
\author{P. G. Martin\altaffilmark{3}}
\author{P. Caselli\altaffilmark{4}}
\author{J. Kauffmann\altaffilmark{5}}
\author{J. E. Pineda\altaffilmark{4}}
\shortauthors{Campbell et al.}

\altaffiltext{1}{Department of Astronomy \& Astrophysics, University of Toronto, 50 St. George St., Toronto, Ontario, Canada, M5S 3H4; \email{jessicalynn.campbell@mail.utoronto.ca}}
\altaffiltext{2}{Dunlap Institute for Astronomy and Astrophysics, University of Toronto, 50 St. George St., Toronto, Ontario, Canada, M5S 3H4}
\altaffiltext{3}{Canadian Institute for Theoretical Astrophysics, University of Toronto, 60 St. George St., Toronto, Ontario, Canada, M5S 3H8}
\altaffiltext{4}{Max-Planck-Institut f{\"u}r extraterrestrische Physik (MPE), Gie{\ss}enbachstrasse 1, D-85741 Garching, Germany}
\altaffiltext{5}{Max-Planck-Institut f{\"u}r Radioastronomie, Auf dem H{\"u}gel 69, 53121 Bonn, Germany}

\begin{abstract}

We report the results of an {\HCOrot} and {\NtDrot} molecular line survey performed toward 91 dense cores in the Perseus molecular cloud using the James Clerk Maxwell Telescope, to identify the fraction of starless and protostellar cores with systematic radial motions. We quantify the {\HCO} asymmetry using a dimensionless asymmetry parameter \deltaV, and identify 20 cores with significant blue or red line asymmetries in optically-thick emission indicative of collapsing or expanding motions, respectively. We separately fit the {\HCO} profiles with an analytic collapse model and determine contraction (expansion) speeds toward 22 cores. Comparing the {\deltaV} and collapse model results, we find that {\deltaV} is a good tracer of core contraction if the optically-thin emission is aligned with the model-derived systemic velocity. The contraction speeds range from subsonic (0.03\,\kms) to supersonic (0.4\,\kms), where the supersonic contraction speeds may trace global rather than local core contraction. Most cores have contraction speeds significantly less than their free-fall speeds. Only 7 of 28 starless cores have spectra well-fit by the collapse model, which more than doubles (15 of 28) for protostellar cores. Starless cores with masses greater than the Jeans mass ($M/M_\mathrm{J} > 1$) are somewhat more likely to show contraction motions. We find no trend of optically-thin non-thermal line width with \MMJ, suggesting that any undetected contraction motions are small and subsonic. Most starless cores in Perseus are either not in a state of collapse or expansion, or are in a very early stage of collapse.

\end{abstract}


\maketitle

\section{Introduction}
\label{sec:intro}

In nearby molecular clouds, stars of low mass like the Sun are found to form in dense cores \citep{Myers1985, Cernicharo1991, Lada1993}. The earliest stage of star formation consists of the gravitationally-driven collapse of a dense core \citep{Larson1969, Shu1977}, forming a protoplanetary disk and a young stellar object (YSO) which accretes surrounding material. The details of the collapse of starless dense cores to their protostellar counterparts remains an active area of research for which the observational study of collapsing dense cores is vital. These regions are predominantly studied with the millimeter and sub-millimeter, probing thermal continuum emission of dust grains \citep[e.g.,][]{Enoch2006} and molecular line emissions of the gas \citep[e.g.,][]{Aikawa2003}, respectively. Infrared observations are also used to characterize embedded protostellar objects \citep[e.g.,][]{Jorgensen2007, Enoch2009, Evans2009}. Observations of molecular transitions that trace high density gas ($n {\,}{\gtrsim}{\,} 10^4$~cm$^{-3}$) are useful probes of the physical and chemical composition of dense cores. In particular, the analysis of optically-thick molecular line emission is an effective tool for analysing the kinematics of core collapse \citep{Myers2000}. Contraction motions in centrally concentrated dense cores produce red-shifted self-absorption in optically-thick line profiles \citep{Leung1977}, resulting in classic blue asymmetries that have become commonly used indicators of contraction motion \citep[see][]{Leung1977, Snell1977, Zhou1992, Mardones1997, Gregersen2000, Myers2000, Lee2011, Chira2014}. \citet{Mardones1997} defined a dimensionless asymmetry parameter to quantify these asymmetries in multi-peaked line profiles, 
\begin{equation} 
\label{eq:deltaV}
{\delta}_v = \frac{ \mathrm{V}_{\mathrm{thick}} - \mathrm{V}_{\mathrm{thin}} }{ {\Delta}\mathrm{V}_{\mathrm{thin}} } \, ,
\end{equation}
where blue asymmetries ({\deltaV} {\textless} 0) identify candidate dense cores undergoing collapse. Here {\Vthick} is the line-of-sight velocity of the brightest component in the optically-thick line profile, {\Vthin} is the line-of-sight velocity of the optically-thin line tracing the systemic velocity of the dense core, and ${\Delta}V_\mathrm{thin}$ is the full width at half maximum (FWHM) of the optically-thin line.

\begin{figure*}[t]
\includegraphics[width=18cm]{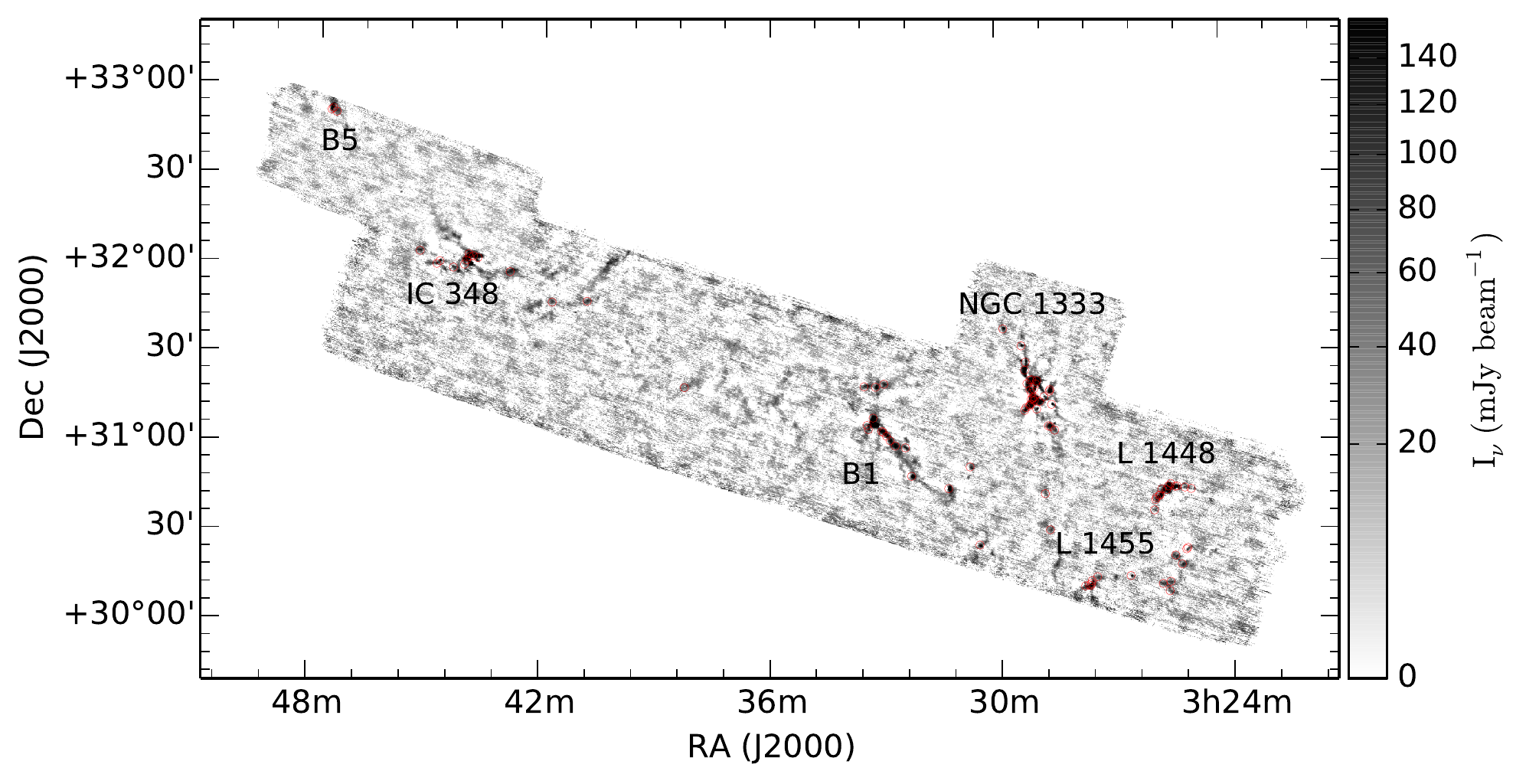}
\caption{Targeted dense cores in our molecular line survey indicated by (red) circles (see Table \ref{table:alltargetsinfo}), overlaid a 1.1-mm continuum map of the Perseus molecular cloud with 31\arcsec~resolution taken with Bolocam on the CSO 10-m telescope, covering 7.5 deg$^2$ (143 pc$^2$ for a distance to Perseus of 250 pc) \citep{Enoch2006}. Six subregions of Perseus are indicated. Brightness scale is as square root. \\
\label{fig:850mapalltargets}}
\end{figure*}

Although contraction motion results in blue asymmetries in optically-thick emission profiles, other systematic gas motions, such as rotation or molecular outflow, could in principle produce similar asymmetries \citep{Snell1980, Adelson1988, Narayanan2002}. Studies of dense cores, however, indicate that rotation has an insignificant effect on core support \citep[][and references therein]{Bergin2007}, suggesting that core contraction and expansion or outflow should dominate over possible line asymmetry effects caused by rotation. Multiple cores along the line of sight can also complicate the observed line profiles, an effect that can be avoided by observing optically-thin molecular transitions to distinguish between the velocity structure of line profiles that are dominated by core kinematics and those resulting from a superposition of cores \citep{Mardones1997, DeVries2005}.

The Perseus molecular cloud (hereafter Perseus) is an example of a moderately clustered star forming region, more complicated than the quiescent, low mass Taurus molecular cloud, but less so than the turbulent Orion molecular cloud currently forming high-mass stars \citep{Ladd1993, Ladd1994, Kirk2007}. Perseus harbors several well-known clustered subregions and is an ideal environment for studying the formation of low-mass stars characterized by moderate levels of clustering and turbulence. Measurements of the distance to Perseus range from 220 to 315 pc, with several authors suggesting that Perseus might be composed of more than one cloud \citep[e.g.,][]{Cernis1990, Cernicharo1991, Cernis1993, Luhman2003}. Given the coherent velocity structure in Perseus and the similar distance estimates across the cloud \citep[e.g.,][]{Schlafly2014}, we will assume that Perseus is a single cloud at a distance of 250~pc.

In this paper we present the results of an {\HCOrot} and {\NtDrot} molecular line survey performed using the James Clerk Maxwell Telescope (JCMT)\footnote{\url{http://www.eaobservatory.org/jcmt/}} toward 91 dense cores in Perseus to identify cores with radial motions and to investigate the relationship between tracers of the contraction (expansion) velocity and core stability. 
We discuss our source selection, observations, and data reduction in Section \ref{sec:obs}.
The line profile fitting and survey results follow in Section \ref{sec:results}.
We present our \edit{core contraction} analysis in Section \ref{sec:analysis}, \edit{where we measure line asymmetries using the asymmetry parameter ({\deltaV}) and determine the core contraction velocities with an analytic model}.
In Section \ref{sec:discussion}, we explore how well {\deltaV} traces core contraction using the results of the analytic model and quantify the gravitational instability of cores using their Jeans mass and non-thermal linewidths, looking for a correlation with core contraction velocity, and discuss the observed contraction speeds of Perseus cores.
A summary is given in Section \ref{sec:summary}.

\section{Observations and Data Reduction}
\label{sec:obs}

\edit{
Table \ref{table:alltargetsinfo} summarizes the 91 dense cores observed, with their Right Ascension (RA) and Declination (Dec) pointing positions. The names in column 1 are as in JCMT proposal ID M07BU33 and the archive of the spectral data.\footnote{\url{http://www.cadc-ccda.hia-iha.nrc-cnrc.gc.ca/en/jcmt/}.}}

\edit{The targets, including both starless and protostellar cores, were first identified in dust continuum surveys, primarily with Bolocam \citep{Enoch2006} using both a peak-finding algorithm and Clumpfind 2D from \citealp{Williams1994}. We call these ``B\#'' and in column 2 is the correspondence with the ``Bolo'' number in the catalog as published (Table~1 in \citealp{Enoch2006}).
Most of the other targets were first identified in JCMT dust maps obtained with SCUBA \citep{Kirk2006} using Clumpfind 2D. We call these targets ``S\#'' where the number is the row number in Table~1 in \citet{Kirk2006}.
We also considered several cold, high column density objects from a list prepared from the far-infrared-based dust map produced by \citet{Schnee2008} (D\#) and one weak submillimeter detection seen in the Bolocam and SCUBA maps that was not included in the published catalogs (W10 in Table~\ref{table:alltargetsinfo} or BS10 in the archive).
}

{\LongTables
\begin{deluxetable*}{lccccccccc}
\tablecolumns{10}
\tablecaption{Summary of Observations}
\tablehead{
\multicolumn{4}{c}{} & \multicolumn{2}{c}{\underline{\HCO}} & \multicolumn{2}{c}{\underline{\NtD}} \\
\colhead{Source} &
\colhead{Bolo} &
\colhead{RA} & 
\colhead{Dec} &
\colhead{$T_\mathrm{peak}$} & 
\colhead{rms} & 
\colhead{$T_\mathrm{peak}$} & 
\colhead{rms} &
\colhead{Protostellar?} & 
\colhead{Reference} \\
\colhead{} &
\colhead{Cross ID} &
\colhead{(J2000)} & 
\colhead{(J2000)} & 
\colhead{(K)} &
\colhead{(K)} &
\colhead{(K)} &
\colhead{(K)} &
\colhead{(Y/N)} & 
\colhead{ID\tablenotemark{e}} \\
\colhead{(1)} &
\colhead{(2)} & 
\colhead{(3)} & 
\colhead{(4)} & 
\colhead{(5)} &
\colhead{(6)} &
\colhead{(7)} & 
\colhead{(8)} &
\colhead{(9)} &
\colhead{(10)} \T
}
\startdata
D26 &\nodata &3$^\mathrm{h}$25$^\mathrm{m}$00$^\mathrm{s}$.3 & 30$\degr$44\arcmin10\farcs5 & \nodata & 0.15 & \nodata & 0.19 & N & \nodata\\ 
B16 & 1 &3$^\mathrm{h}$25$^\mathrm{m}$07$^\mathrm{s}$.8 & 30$\degr$24\arcmin21\farcs6 & 1.48 & 0.16 & \nodata & 0.14 & N & \nodata \\ 
B15\tablenotemark{d} & 2 & 3$^\mathrm{h}$25$^\mathrm{m}$09$^\mathrm{s}$.7 & 30$\degr$23\arcmin53\farcs0 & 1.73 & 0.21 & \nodata & 0.16 & Y & P \\ 
B31 & 3 & 3$^\mathrm{h}$25$^\mathrm{m}$10$^\mathrm{s}$.1 & 30$\degr$44\arcmin40\farcs9 & 1.58 & 0.22 & \nodata & 0.24 & N & \nodata \\ 
B11 & 4 & 3$^\mathrm{h}$25$^\mathrm{m}$17$^\mathrm{s}$.1 & 30$\degr$18\arcmin52\farcs9 & 1.18 & 0.20 & \nodata & 0.15 & N & \nodata \\ 
B30 & 5 & 3$^\mathrm{h}$25$^\mathrm{m}$22$^\mathrm{s}$.5 & 30$\degr$45\arcmin06\farcs6 & 1.46 & 0.23 & \nodata & 0.15 & Y & J1 \\ 
S57\tablenotemark{d,f} &\nodata& 3$^\mathrm{h}$25$^\mathrm{m}$25$^\mathrm{s}$.7 & 30$\degr$45\arcmin01\farcs7 & 2.90 & 0.29 & 0.86 & 0.15 & Y & C \\ 
B13 & 6 & 3$^\mathrm{h}$25$^\mathrm{m}$26$^\mathrm{s}$.9 & 30$\degr$21\arcmin52\farcs7 & \nodata & 0.16 & \nodata & 0.22 & N & \nodata \\ 
D38 &\nodata& 3$^\mathrm{h}$25$^\mathrm{m}$32$^\mathrm{s}$.3 & 30$\degr$46\arcmin00\farcs3 & 2.60 & 0.35 & \nodata & 0.17 & N & \nodata \\ 
B6 & 7 & 3$^\mathrm{h}$25$^\mathrm{m}$35$^\mathrm{s}$.5 & 30$\degr$13\arcmin06\farcs2 & \nodata & 0.18 & \nodata & 0.25 & N & \nodata \\ 
B29\tablenotemark{b,c} & 8 & 3$^\mathrm{h}$25$^\mathrm{m}$36$^\mathrm{s}$.0 & 30$\degr$45\arcmin10\farcs8 & 4.52 & 0.25 & 1.60 & 0.07 & Y & J2 \\ 
B1 & 9 & 3$^\mathrm{h}$25$^\mathrm{m}$37$^\mathrm{s}$.2 & 30$\degr$09\arcmin55\farcs3 & \nodata & 0.19 & \nodata & 0.28 & N & \nodata \\ 
B25\tablenotemark{a,c} & 10 & 3$^\mathrm{h}$25$^\mathrm{m}$38$^\mathrm{s}$.9 & 30$\degr$43\arcmin59\farcs7 & 5.01 & 0.19 & 1.09 & 0.19 & Y & J3 \\ 
B27 & 11 & 3$^\mathrm{h}$25$^\mathrm{m}$46$^\mathrm{s}$.3 & 30$\degr$44\arcmin14\farcs0 & 2.06 & 0.22 & \nodata & 0.17 & N & \nodata \\ 
B4 & 12 & 3$^\mathrm{h}$25$^\mathrm{m}$47$^\mathrm{s}$.5 & 30$\degr$12\arcmin26\farcs4 & \nodata & 0.24 & \nodata & 0.24 & N & \nodata \\ 
B24 & 13 & 3$^\mathrm{h}$25$^\mathrm{m}$49$^\mathrm{s}$.3 & 30$\degr$42\arcmin15\farcs2 & 1.43 & 0.27 & 0.80 & 0.16 & N & \nodata \\ 
B23 & 14 & 3$^\mathrm{h}$25$^\mathrm{m}$50$^\mathrm{s}$.6 & 30$\degr$42\arcmin01\farcs7 & 1.68 & 0.29 & 0.81 & 0.16 & N & \nodata \\ 
B22 & 15 & 3$^\mathrm{h}$25$^\mathrm{m}$55$^\mathrm{s}$.1 & 30$\degr$41\arcmin26\farcs2 & \nodata & 0.16 & \nodata & 0.22 & N & \nodata \\ 
B21 & 16 & 3$^\mathrm{h}$25$^\mathrm{m}$56$^\mathrm{s}$.4 & 30$\degr$40\arcmin42\farcs9 & \nodata & 0.20 & \nodata & 0.28 & N & \nodata \\ 
B20 & 17 & 3$^\mathrm{h}$25$^\mathrm{m}$58$^\mathrm{s}$.5 & 30$\degr$37\arcmin13\farcs5 & 1.37 & 0.21 & \nodata & 0.20 & N & \nodata \\ 
B10 & 18 & 3$^\mathrm{h}$26$^\mathrm{m}$37$^\mathrm{s}$.2 & 30$\degr$15\arcmin18\farcs9 & \nodata & 0.32 & \nodata & 0.12 & Y & J4 \\ 
B8 & 20 & 3$^\mathrm{h}$27$^\mathrm{m}$29$^\mathrm{s}$.5 & 30$\degr$15\arcmin09\farcs1 & \nodata & 0.30 & \nodata & 0.20 & N & \nodata \\ 
B7\tablenotemark{a} & 21 & 3$^\mathrm{h}$27$^\mathrm{m}$37$^\mathrm{s}$.9 & 30$\degr$13\arcmin53\farcs2 & 1.84 & 0.18 & \nodata & 0.14 & Y & J4 \\ 
B5\tablenotemark{c} & 22 & 3$^\mathrm{h}$27$^\mathrm{m}$39$^\mathrm{s}$.0 & 30$\degr$12\arcmin53\farcs5 & 4.08 & 0.22 & \nodata & 0.14 & Y & J6 \\ 
S50 &\nodata& 3$^\mathrm{h}$27$^\mathrm{m}$40$^\mathrm{s}$.0 & 30$\degr$12\arcmin12\farcs8 & 1.40 & 0.10 & \nodata & 0.22 & N & \nodata \\ 
B2 & 23 & 3$^\mathrm{h}$27$^\mathrm{m}$42$^\mathrm{s}$.7 & 30$\degr$12\arcmin24\farcs5 & 3.09 & 0.19 & 0.76 & 0.13 & Y & J7 \\ 
B3 &24 & 3$^\mathrm{h}$27$^\mathrm{m}$48$^\mathrm{s}$.3 & 30$\degr$12\arcmin08\farcs2 & 1.91 & 0.22 & \nodata & 0.13 & Y & J8 \\ 
B43 & 26 & 3$^\mathrm{h}$28$^\mathrm{m}$32$^\mathrm{s}$.4 & 31$\degr$04\arcmin43\farcs4 & 1.74 & 0.10 & \nodata & 0.14 & N & \nodata \\ 
B59 & 29 & 3$^\mathrm{h}$28$^\mathrm{m}$36$^\mathrm{s}$.7 & 31$\degr$13\arcmin23\farcs8 & 4.54 & 0.31 & \nodata & 0.13 & Y & J11 \\ 
B45 & 30 & 3$^\mathrm{h}$28$^\mathrm{m}$38$^\mathrm{s}$.8 & 31$\degr$05\arcmin54\farcs2 & 2.58 & 0.22 & \nodata & 0.19 & Y & J12 \\ 
S46\tablenotemark{c} &\nodata& 3$^\mathrm{h}$28$^\mathrm{m}$39$^\mathrm{s}$.1 & 31$\degr$18\arcmin24\farcs1 & 1.80 & 0.23 & \nodata & 0.36 & Y & D2376 \\ 
S44\tablenotemark{c} &\nodata& 3$^\mathrm{h}$28$^\mathrm{m}$40$^\mathrm{s}$.1 & 31$\degr$17\arcmin48\farcs6 & 1.97 & 0.28 & \nodata & 0.19 & Y & J13 \\ 
B19 & 32 & 3$^\mathrm{h}$28$^\mathrm{m}$41$^\mathrm{s}$.7 & 30$\degr$31$\degr$12\farcs3 & \nodata & 0.16 & \nodata & 0.20 & N & \nodata \\ 
B46\tablenotemark{b} & 33 & 3$^\mathrm{h}$28$^\mathrm{m}$42$^\mathrm{s}$.5 & 31$\degr$06\arcmin13\farcs1 & 1.82 & 0.30 & 0.69 & 0.11 & N & \nodata \\ 
B64 & 35 & 3$^\mathrm{h}$28$^\mathrm{m}$48$^\mathrm{s}$.5 & 31$\degr$16\arcmin03\farcs0 & 1.77 & 0.11 & \nodata & 0.14 & Y & D2388 \\ 
B26 & 36 & 3$^\mathrm{h}$28$^\mathrm{m}$48$^\mathrm{s}$.8 & 30$\degr$43\arcmin24\farcs5 & \nodata & 0.10 & \nodata & 0.32 & N & \nodata \\ 
B60\tablenotemark{a,c} & 38 & 3$^\mathrm{h}$28$^\mathrm{m}$55$^\mathrm{s}$.3 & 31$\degr$14\arcmin27\farcs8 & 3.99 & 0.20 & \nodata & 0.22 & Y & J15 \\ 
B80\tablenotemark{c,f} & 40 & 3$^\mathrm{h}$28$^\mathrm{m}$59$^\mathrm{s}$.5 & 31$\degr$21\arcmin28\farcs7 & 4.72 & 0.22 & \nodata & 0.25 & Y & J17 \\ 
B56 & 41 & 3$^\mathrm{h}$29$^\mathrm{m}$00$^\mathrm{s}$.2 & 31$\degr$11\arcmin52\farcs9 & 3.08 & 0.23 & \nodata & 0.37 & Y & J18 \\ 
B72\tablenotemark{c} & 42 & 3$^\mathrm{h}$29$^\mathrm{m}$01$^\mathrm{s}$.4 & 31$\degr$20\arcmin23\farcs1 & 10.69 & 0.23 & 0.89 & 0.15 & Y & J19 \\ 
S39 &\nodata& 3$^\mathrm{h}$29$^\mathrm{m}$03$^\mathrm{s}$.2 & 31$\degr$15\arcmin53\farcs7 & 6.92 & 0.16 & 1.36 & 0.16 & Y & J20 \\ 
S38\tablenotemark{a} &\nodata& 3$^\mathrm{h}$29$^\mathrm{m}$03$^\mathrm{s}$.7 & 31$\degr$14\arcmin47\farcs8 & 5.66 & 0.28 & \nodata & 0.22 & Y & J21 \\ 
B71 & 44 & 3$^\mathrm{h}$29$^\mathrm{m}$04$^\mathrm{s}$.5 & 31$\degr$18\arcmin42\farcs7 & 2.35 & 0.10 & \nodata & 0.15 & N & \nodata \\ 
S37 &\nodata& 3$^\mathrm{h}$29$^\mathrm{m}$06$^\mathrm{s}$.5 & 31$\degr$15\arcmin36\farcs3 & 6.35 & 0.20 & 1.38 & 0.18 & N & \nodata \\ 
S35\tablenotemark{c} &\nodata& 3$^\mathrm{h}$29$^\mathrm{m}$07$^\mathrm{s}$.4 & 31$\degr$21\arcmin48\farcs4 & 3.53 & 0.23 & \nodata & 0.34 & Y & E50 \\ 
B65\tablenotemark{d} & 45 & 3$^\mathrm{h}$29$^\mathrm{m}$07$^\mathrm{s}$.8 & 31$\degr$17\arcmin18\farcs7 & 2.06 & 0.09 & \nodata & 0.23 & Y & S \\ 
S34 &\nodata& 3$^\mathrm{h}$29$^\mathrm{m}$08$^\mathrm{s}$.8 & 31$\degr$15\arcmin12\farcs8 & 4.19 & 0.15 & 1.47 & 0.19 & N & \nodata \\ 
S33\tablenotemark{a,b,c} &\nodata& 3$^\mathrm{h}$29$^\mathrm{m}$09$^\mathrm{s}$.9 & 31$\degr$13\arcmin31\farcs2 & 6.41 & 0.08 & 1.23 & 0.15 & Y & J22 \\ 
S32\tablenotemark{c} &\nodata& 3$^\mathrm{h}$29$^\mathrm{m}$10$^\mathrm{s}$.2 & 31$\degr$21\arcmin42\farcs9 & 6.27 & 0.21 & \nodata & 0.18 & Y & E50 \\ 
B67\tablenotemark{c} & 49 & 3$^\mathrm{h}$29$^\mathrm{m}$10$^\mathrm{s}$.5 & 31$\degr$18\arcmin25\farcs0 & 3.85 & 0.19 & \nodata & 0.19 & Y & J23 \\ 
S30\tablenotemark{a,c} &\nodata& 3$^\mathrm{h}$29$^\mathrm{m}$11$^\mathrm{s}$.3 & 31$\degr$13\arcmin07\farcs5 & 5.05 & 0.16 & 2.06 & 0.19 & Y & J25 \\ 
B74 & 50 & 3$^\mathrm{h}$29$^\mathrm{m}$15$^\mathrm{s}$.0 & 31$\degr$20\arcmin32\farcs1 & 5.77 & 0.26 & \nodata & 0.17 & N & \nodata \\ 
B55 & 51 & 3$^\mathrm{h}$29$^\mathrm{m}$17$^\mathrm{s}$.0 & 31$\degr$12\arcmin25\farcs8 & 1.85 & 0.17 & 0.96 & 0.19 & N & \nodata \\ 
B83\tablenotemark{b,c} & 52 & 3$^\mathrm{h}$29$^\mathrm{m}$17$^\mathrm{s}$.2 & 31$\degr$27\arcmin44\farcs4 & 4.34 & 0.20 & 0.44 & 0.12 & Y & J27 \\ 
B82\tablenotemark{a} & 53 & 3$^\mathrm{h}$29$^\mathrm{m}$18$^\mathrm{s}$.4 & 31$\degr$25\arcmin02\farcs7 & 3.59 & 0.28 & 0.99 & 0.18 & N & \nodata \\ 
B54 & 55 & 3$^\mathrm{h}$29$^\mathrm{m}$19$^\mathrm{s}$.1 & 31$\degr$11\arcmin32\farcs1 & 1.81 & 0.16 & 0.44\tablenotemark{a} & 0.19 & N & \nodata \\ 
B85\tablenotemark{b} & 57 & 3$^\mathrm{h}$29$^\mathrm{m}$23$^\mathrm{s}$.4 & 31$\degr$33\arcmin15\farcs7 & 2.63 & 0.23 & 0.72 & 0.13 & Y & J30 \\ 
B87 & 59 & 3$^\mathrm{h}$29$^\mathrm{m}$52$^\mathrm{s}$.0 & 31$\degr$39\arcmin03\farcs4 & 4.91 & 0.26 & \nodata & 0.21 & Y & J31 \\ 
B17 & 62 & 3$^\mathrm{h}$30$^\mathrm{m}$32$^\mathrm{s}$.0 & 30$\degr$26\arcmin24\farcs0 & 1.82 & 0.23 & 0.57 & 0.13 & Y & E7 \\ 
B34 & 63 & 3$^\mathrm{h}$30$^\mathrm{m}$46$^\mathrm{s}$.1 & 30$\degr$52\arcmin44\farcs0 & 1.24 & 0.21 & \nodata & 0.36 & N & \nodata \\ 
B28 & 65 & 3$^\mathrm{h}$31$^\mathrm{m}$20$^\mathrm{s}$.0 & 30$\degr$45\arcmin30\farcs5 & 2.46 & 0.18 & \nodata & 0.21 & Y & J32 \\ 
B33 & 66 & 3$^\mathrm{h}$32$^\mathrm{m}$18$^\mathrm{s}$.0 & 30$\degr$49\arcmin45\farcs4 & 3.10 & 0.20 & 1.29 & 0.18 & Y & J33 \\
B36 & 67 & 3$^\mathrm{h}$32$^\mathrm{m}$27$^\mathrm{s}$.4 & 30$\degr$59\arcmin22\farcs0 & \nodata & 0.31 & \nodata & 0.16 & N & \nodata\\ 
B37 & 70 & 3$^\mathrm{h}$32$^\mathrm{m}$43$^\mathrm{s}$.2 & 31$\degr$00\arcmin00\farcs0 & 1.25 & 0.22 & \nodata & 0.17 & N & \nodata\\ 
B38 & 71 & 3$^\mathrm{h}$32$^\mathrm{m}$51$^\mathrm{s}$.3 & 31$\degr$01\arcmin47\farcs9 & 0.43 & 0.10 & \nodata & 0.19 & N & \nodata \\ 
B40 & 72 & 3$^\mathrm{h}$32$^\mathrm{m}$57$^\mathrm{s}$.0 & 31$\degr$03\arcmin21\farcs1 & 0.64 & 0.15 & \nodata & 0.21 & N & \nodata \\ 
B75 & 73 & 3$^\mathrm{h}$33$^\mathrm{m}$00$^\mathrm{s}$.6 & 31$\degr$20\arcmin50\farcs0 & \nodata & 0.21 & \nodata & 0.12 & N & \nodata \\ 
B41 & 74 & 3$^\mathrm{h}$33$^\mathrm{m}$02$^\mathrm{s}$.0 & 31$\degr$04\arcmin33\farcs5 & 0.92 & 0.10 & \nodata & 0.15 & N & \nodata \\ 
B42 & 75 & 3$^\mathrm{h}$33$^\mathrm{m}$04$^\mathrm{s}$.3 & 31$\degr$04\arcmin57\farcs5 & 1.31 & 0.16 & \nodata & 0.18 & N & \nodata \\ 
B73 & 78 & 3$^\mathrm{h}$33$^\mathrm{m}$13$^\mathrm{s}$.8 & 31$\degr$19\arcmin51\farcs3 & \nodata & 0.21 & \nodata & 0.17 & Y & J34 \\ 
B52\tablenotemark{b} & 80 & 3$^\mathrm{h}$33$^\mathrm{m}$17$^\mathrm{s}$.9 & 31$\degr$09\arcmin27\farcs8 & 6.93 & 0.23 & 1.32 & 0.13 & Y & J38 \\ 
B44 & 82 & 3$^\mathrm{h}$33$^\mathrm{m}$25$^\mathrm{s}$.2 & 31$\degr$05\arcmin35\farcs4 & 0.52 & 0.10 & \nodata & 0.16 & N & \nodata \\ 
B48 & 84 & 3$^\mathrm{h}$33$^\mathrm{m}$27$^\mathrm{s}$.3 & 31$\degr$06\arcmin58\farcs7 & 2.45 & 0.22 & \nodata & 0.18 & Y & J40 \\ 
B77 & 85 & 3$^\mathrm{h}$33$^\mathrm{m}$31$^\mathrm{s}$.8 & 31$\degr$20\arcmin01\farcs9 & 0.92 & 0.19 & \nodata & 0.21 & N & \nodata \\ 
W10 &\nodata& 3$^\mathrm{h}$38$^\mathrm{m}$15$^\mathrm{s}$.1 & 31$\degr$19\arcmin44\farcs7 & \nodata & 0.36 & \nodata & 0.37 & N & \nodata \\ 
B91 & 89 & 3$^\mathrm{h}$40$^\mathrm{m}$49$^\mathrm{s}$.7 & 31$\degr$48\arcmin34\farcs2 & \nodata & 0.27 & \nodata & 0.21 & N & \nodata \\ 
B92 & 93 & 3$^\mathrm{h}$41$^\mathrm{m}$45$^\mathrm{s}$.2 & 31$\degr$48\arcmin08\farcs7 & \nodata & 0.72 & \nodata & \nodata & N & \nodata \\ 
B95 & 97 & 3$^\mathrm{h}$42$^\mathrm{m}$52$^\mathrm{s}$.5 & 31$\degr$58\arcmin11\farcs6 & \nodata & 0.19 & \nodata & 0.20 & N & \nodata \\ 
S13\tablenotemark{a} &\nodata& 3$^\mathrm{h}$43$^\mathrm{m}$44$^\mathrm{s}$.0 & 32$\degr$02\arcmin46\farcs6 & 2.09 & 0.26 & \nodata & 0.35 & N & \nodata \\ 
B112\tablenotemark{c} & 102 & 3$^\mathrm{h}$43$^\mathrm{m}$51$^\mathrm{s}$.1 & 32$\degr$03\arcmin21\farcs1 & 4.25 & 0.25 & 0.94 & 0.13 & Y & J41 \\ 
B113\tablenotemark{b,c} & 104 & 3$^\mathrm{h}$43$^\mathrm{m}$57$^\mathrm{s}$.2 & 32$\degr$03\arcmin01\farcs6 & 6.56 & 0.51 & 0.53 & 0.15 & Y & J43 \\ 
B115 & 105 & 3$^\mathrm{h}$43$^\mathrm{m}$58$^\mathrm{s}$.2 & 32$\degr$04\arcmin01\farcs3 & 2.70 & 0.22 & \nodata & 0.15 & N & \nodata \\ 
B106\tablenotemark{c} & 106 & 3$^\mathrm{h}$44$^\mathrm{m}$01$^\mathrm{s}$.8 & 32$\degr$01\arcmin54\farcs6 & 6.18 & 0.33 & 0.97 & 0.16 & Y & J47 \\ 
B102\tablenotemark{c} & 109 & 3$^\mathrm{h}$44$^\mathrm{m}$05$^\mathrm{s}$.1 & 32$\degr$00\arcmin27\farcs8 & 0.79 & 0.11 & \nodata & 0.14 & Y & D2608 \\ 
B100 & 113 & 3$^\mathrm{h}$44$^\mathrm{m}$22$^\mathrm{s}$.6 & 31$\degr$59\arcmin23\farcs7 & 1.65 & 0.10 & \nodata & 0.15 & Y & E61 \\ 
B104\tablenotemark{c} & 116 & 3$^\mathrm{h}$44$^\mathrm{m}$44$^\mathrm{s}$.2 & 32$\degr$01\arcmin26\farcs9 & 4.24 & 1.21 & \nodata & 0.34 & Y & J48 \\ 
B103 & 117 & 3$^\mathrm{h}$44$^\mathrm{m}$48$^\mathrm{s}$.9 & 32$\degr$00\arcmin31\farcs8 & 1.86 & 0.19 & \nodata & 0.14 & N & \nodata \\ 
B116 & 119 & 3$^\mathrm{h}$45$^\mathrm{m}$15$^\mathrm{s}$.9 & 32$\degr$04\arcmin48\farcs7 & 1.70 & 0.11 & \nodata & 0.14 & N & \nodata \\ 
B121 & 121& 3$^\mathrm{h}$47$^\mathrm{m}$33$^\mathrm{s}$.5 & 32$\degr$50\arcmin55\farcs1 & \nodata & 0.18 & \nodata & 0.19 & N & \nodata \\ 
S2 &\nodata& 3$^\mathrm{h}$47$^\mathrm{m}$39$^\mathrm{s}$.0 & 32$\degr$52\arcmin11\farcs2 & 1.54 & 0.20 & \nodata & 0.23 & N & \nodata \\ 
S1 &\nodata& 3$^\mathrm{h}$47$^\mathrm{m}$41$^\mathrm{s}$.8 & 32$\degr$51\arcmin40\farcs3 & 4.39 & 0.18 & \nodata & 0.27 & Y & J49 
\enddata
\tablenotetext{a}{More than one dense core along the line of sight (Section \edit{\ref{sec:analysis}}).}
\tablenotetext{b}{Systemic velocity sufficiently aligned with brightest \HCO~peak (Section \edit{\ref{sec:analysis}}).}
\tablenotetext{c}{More than one nearby ({\textless} 40{\arcsec}) YSO (Section \ref{sec:obs}).} 
\tablenotetext{d}{Classified as protostellar based on the detection of associated outflows, despite lacking infrared point-source detections.} 
\tablenotetext{e}{Letters of the YSO reference ID designating their catalog source -- J for \citet{Jorgensen2007}, E for \citet{Enoch2009}, D for \citet{Dunham2015}, C for \citet{Chen2010}, P for \citet{Pineda2011}, S for \citet{Schnee2012} -- and associated source number in that catalog where applicable.}
\tablenotetext{f}{{\HCOrot} line profiles too complex for the analysis in this paper.}
\label{table:alltargetsinfo}
\end{deluxetable*}
}

\edit{To optimize the use of telescope time within the allocation our initial list contained about 110 targets concentrating on pointings with
the brightest emission in the 1.1~mm Bolocam map (\citealp{Enoch2006}, reproduced in Figure~\ref{fig:850mapalltargets}). As a result, the targets observed had 
$I_\nu~{\geq}~110 \;\mathrm{mJy\, beam}^{-1}$ at 1.1~mm 
or, using the conversions adopted by \citet{Enoch2006}, extinction $A_{\mathrm{V}} ~{\geq}~7.7 \; \mathrm{mag}$ and H$_2$ column density N(H$_2)~{\geq}~7.3{\times}10^{21}{\;}\mathrm{cm}^{-2}$. The candidates were ranked in three groups: (i) in L1448, (ii) those with a position match with a counterpart in \citet{Kirk2007} within 15\arcsec, and (iii) the rest; priorities within these groups were randomized. In the end data were obtained for 91 targets. As seen in Figure~\ref{fig:850mapalltargets}, where circles indicate the observed dense core positions, the cores surveyed span a significant extent of the projected cloud in an unbiased way.}

We identified protostellar cores as those within 40{\arcsec} (approximately the radius of typical Perseus cores) of protostellar sources in several catalogs as noted in the table \citep{Jorgensen2007, Enoch2009, Dunham2015}. We classified three more cores as protostellar based on the detection of associated outflows, despite their lacking infrared point-source detections \citep{Chen2010, Pineda2011, Schnee2012} (specified in notes to Table~\ref{table:alltargetsinfo}). Cores found to have more than one associated YSO are specified in notes to Table \ref{table:alltargetsinfo}; for these we reference the closest one if multiple exist in the same catalog, otherwise we reference the first occurrence. Our core sample is divided roughly evenly into protostellar (43 of 91, or 47\,\%) and starless (48 of 91, or 53\,\%) cores.

Pointed observations of the Perseus cores were performed using the JCMT. Targets were observed in the {\HCOrot} and {\NtDrot} rotational transitions in position-switching mode, with assumed rest frequencies of 267.557619\,GHz and 231.321665\,GHz, respectively. Observations were taken with an 1800 MHz bandwidth using the A3 front-end receiver and ACSIS back-end. The spectral resolution was 30.5 kHz, corresponding to a velocity resolution of 0.03 \kms~for {\HCOrot} and 0.04 \kms~for {\NtDrot}. The JCMT's half-power beam width (HPBW) at 230\,GHz subtends 19\farcs7, which corresponds to a physical size of 0.02\,pc at the assumed distance of 250\,pc to Perseus. This is roughly four times smaller than the typical size of dense cores in Perseus \citep{Jijina1999, Kirk2006, Rosolowsky2008}. Our molecular line observations are thus pencil-beam pointings through each dense core. Observations were conducted between September 2007 and September 2009. 

We found a discrepancy between the line-of-sight local-standard-of-rest velocities (\VLSR) between the detected sources in the archival JCMT data and the known values of the Perseus cores, where the observed {\VLSR} values were significantly higher than expected. We corrected this offset by using the center frequencies in the archival data headers to convert from frequency to velocity, rather than the observed rest frequencies of {\HCO} and {\NtD}. This produced spectra that agreed with the known {\VLSR} values. As a check, we obtained the raw data for several targets and re-reduced these using the Starlink software (Hikianalia release). The reduced velocity axes agreed with our correction to the calibrated archival data, and consequently with the known core {\VLSR} values. All spectra were calibrated from antenna temperature ({\Atemp}) to main-beam temperature ({\TMB}) using a main-beam efficiency of ${\eta}_\mathrm{MB}=0.69$. 

\begin{figure}[h]
\begin{center}
\includegraphics[width=8cm]{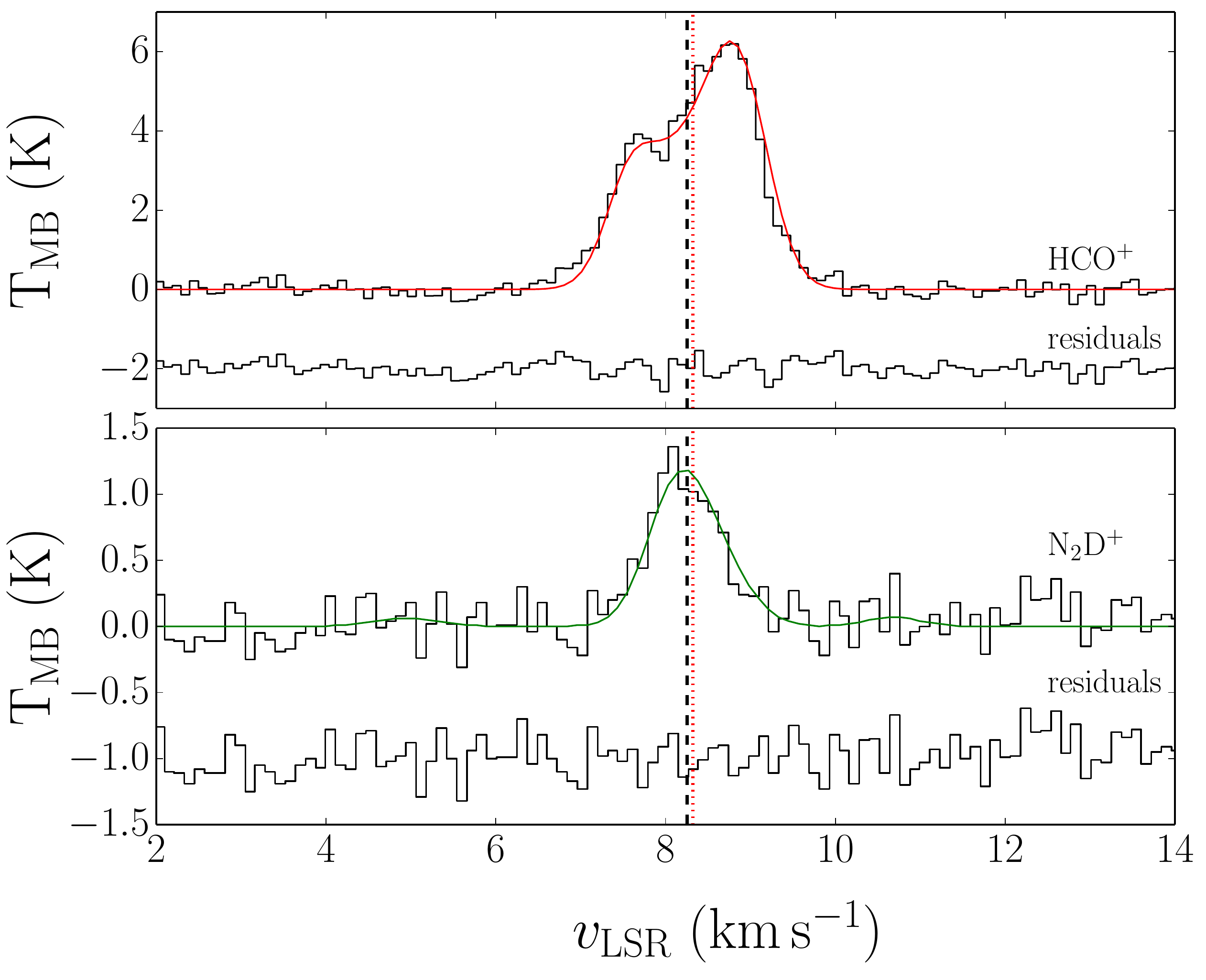}
\caption{Top: The optically-thick {\HCO} emission profile for source S39 after removal of a broad Gaussian (see text) over-plotted by the best fit of the {\Hill} model (red), with the resulting residuals shown below. In both plots, the observed systemic velocity ({\Vthin}) and the modeled value ({\Vmod}) are shown in dashed black and dotted red vertical lines, respectively.
Bottom: The optically-thin {\NtD} emission profile for source S39 over-plotted by the hyperfine fit (green), with the resulting residuals shown below.
Note the different scales in {\TMB} between the two plots.} 
\label{fig:examplespectra}
\end{center}
\end{figure}

Targets observed more than once were co-added by taking their weighted average, where each spectrum was weighted by the inverse square of the baseline rms. Spectra were then smoothed and decimated to a velocity resolution of 0.1 \kms. This velocity resolution ensures that three independent velocity channels span the narrowest emission line profile. As listed in Table~\ref{table:alltargetsinfo}, the resulting baseline rms for {\HCO} ranged from 0.08 K to 0.72 K, and for {\NtD} from 0.07 K to 0.37~K. Spectral line fitting was performed using the Python Scipy {\tt curve\_fit} routine.\footnote{\url{http://www.scipy.org/}} Linear baselines were fit to each spectrum near emission lines for baseline subtraction. A linear correction to the baseline was insufficient for one target in {\HCO} (B60) and so a third degree polynomial was used. 

\section{\edit{Line Profile Fitting and Detection Rates}}
\label{sec:results}

In the Appendix all {\HCO} line profiles are shown in Figure \ref{fig:allhcoprofiles} and all {\NtD} line profiles are shown in Figure \ref{fig:alln2dprofiles}, with sources in order of increasing RA as in Table~\ref{table:alltargetsinfo}.

We model the {\NtD} hyperfine structure following \citet{Friesen2010} using the hyperfine component frequencies and relative line strengths from \citet{Gerin2001}. Even when the individual hyperfine components of {\NtDrot} are unresolved, the hyperfine structure combines to broaden and make asymmetric the line profile which can result in overestimated line widths and inaccurate source velocities if not properly modeled. Due to the low S/N of the {\NtD} emission lines we fixed the excitation temperature to 7\,K, the mean value found for {\NtH} for the Perseus cores \citep{Kirk2007}. The results are shown in Columns 2 and 3 of Table \ref{table:kinematics}, where {\Vthin} is the systemic velocity and {\sigmathin} is the velocity dispersion. 

\edit{For cores not detected in {\NtD}, we obtained optically-thin measurements from, in order of precedence, {\NtH} \citep{Kirk2007} and {\ammonia} \citep{Rosolowsky2008} line surveys, 
requiring a positional match better than 15\arcsec.
For such cores with {\NtH} or {\ammonia} measurements, we list the reference ID in Column 4 of Table \ref{table:kinematics}.
While {\NtH} has a similar critical density to that of {\NtD} \citep{Tafalla2002}, {\ammonia} has a slightly lower critical density \citep{Swade1989}, which might result in tracing somewhat different material. Where the cores in our contraction analysis have both {\NtD} detections and {\NtH} or {\ammonia} detections, we compared the observed systemic velocities, and also line widths, measured with the different molecular tracers. We found good agreement: the average difference in systemic velocity and line width between {\NtD} and {\NtH} is $-0.002$\,\kms\, and $0.12$\,\kms\, ($31\,\%$), respectively. Similarly for {\NtD} and {\ammonia} the average differences are $0.02$\,\kms\, and $0.09$\,\kms\, ($26\,\%$), respectively.}

For {\HCO} emission profiles, we fit a linear combination of Gaussian components to the emission lines to measure the source velocity, amplitude, and velocity dispersion. The {\HCO} profiles often showed complex velocity structure and so were fit with a combination of Gaussians. An F-test was used to determine the number of Gaussians required to capture the full extent of the velocity structure, with a false-rejection probability of 0.05 corresponding to a confidence level of 95\,\%. To be fit accurately, most {\HCO} profiles required only one or two Gaussian components, with some requiring a third Gaussian to fit broad, high velocity line wings likely from YSO-driven molecular outflows. One source (S33) was unique in that a fourth Gaussian component was required to fit an additional feature. The statistics from the Gaussian profile fitting results are summarized in Table \ref{table:summary}.

We set a signal-to-noise ratio (S/N) threshold of 3 in {\TMB} using the (brightest) Gaussian fit amplitude for a source detection. Of the 91 observed sources, 72 (79\,\%) yield detections in {\HCO}. Protostellar sources were more likely to be detected (41 of 43 cores, or 95\,\%) than their starless counterparts (31 of 48 cores, or 65\,\%). The observed {\HCO} line profiles varied from symmetric and of approximately Gaussian shape to very complex. We omit two source detections in {\HCO} (B80, S57) having line profiles too complex for the analysis in this paper. We detect 23 (25\,\%) of the dense cores in {\NtD}, all of which were detected in their {\HCO} counterparts. We additionally include source B54; it exhibits an {\NtD} detection with S/N of only 2.3 but the emission profile is well aligned with its {\HCO} counterpart, indicating that it is likely a true detection arising from the same source rather than a false-detection. For detected molecular species, we list the maximum amplitude in {\TMB} of the data as $T_\mathrm{peak}$ in Columns 4 and 6 of Table \ref{table:alltargetsinfo}. For all sources, we list the baseline rms in Columns 5 and 7. 

We show a representative optically-thick {\HCO} emission profile (top) with its corresponding optically-thin {\NtD} counterpart (bottom) for source S39 in Figure \ref{fig:examplespectra}. 

\begin{deluxetable*}{lcccccccccccc}
\tablecolumns{11}
\tablecaption{Summary of Perseus Dense Core Kinematics}
\tablehead{
\colhead{} & \multicolumn{2}{c}{{\NtD}/{\NtH}/{\ammonia} } & \colhead{} & \multicolumn{6}{c}{\HCO~HILL5} & \colhead{} & \multicolumn{2}{c}{Asymmetry} \\
\colhead{} & \multicolumn{2}{c}{Fit Results} & \colhead{} & \multicolumn{6}{c}{Fit Results} & \colhead{} & \multicolumn{2}{c}{Analysis} \\
\cline{2-3} \cline{5-10} \cline{12-13} \\
\colhead{Source} &
\colhead{\Vthin} &
\colhead{{\sigmathin}} &
\colhead{Reference} &
\colhead{\Vmod} &
\colhead{$\sigma$} &
\colhead{\Vin} &
\colhead{\Tex} &
\colhead{$\tau$} &
\colhead{\redchisq} &
\colhead{} &
\colhead{\deltaV} & 
\colhead{\deltaVstar} \\
\colhead{} &
\colhead{(\kms)} &
\colhead{(\kms)} &
\colhead{ID\tablenotemark{a}} &
\colhead{(\kms)} &
\colhead{(\kms)} &
\colhead{(\kms)} &
\colhead{(K)} &
\colhead{} &
\colhead{} &
\colhead{} & 
\colhead{} &
\colhead{} \\
\colhead{(1)} &
\colhead{(2)} & 
\colhead{(3)} & 
\colhead{(4)} & 
\colhead{(5)} &
\colhead{(6)} & 
\colhead{(7)} &
\colhead{(8)} &
\colhead{(9)} &
\colhead{(10)} & 
\colhead{} &
\colhead{(11)} &
\colhead{(12)} \T
}
\startdata
B16 & 4.1360(1) & 0.1150(1) & R3 & 4.06(2) & 0.13(1) & $-0.11(4)$ & 7.1(4) & 3.4(8) & 1.0 & & \nodata & \nodata\T\\
B15 & 3.9490(2) & 0.1550(2) & R4 & 4.11(3) & 0.14(2) & 0.26(3) & 6.8(3) & 4(1) & 0.9 & & \nodata & \nodata \\ 
B30 & 4.1350(1) & 0.1680(1) & R7 & 4.12(3) & 0.36(2) & $-0.06(4)$ & 7.8(3) & 4.5(8) & 1.2 & & \nodata & \nodata \\ 
B20 & 3.5590(9) & 0.210(1) & R21 & \nodata & \nodata & \nodata & \nodata & \nodata & \nodata & & 0.42(7) & \nodata \\ 
B5 & 4.720(5) & 0.263(2) & K135 & \nodata & \nodata & \nodata & \nodata & \nodata & \nodata & & $-0.69(8)$ & \nodata \\ 
S50 & 4.9280(3) & 0.1860(3) & R33 & 5.11(2) & 0.29(8) & 0.3(2) & 7.6(7) & 2(2) & 1.6 & & \nodata & \nodata \\ 
B3 & 5.060(7) & 0.234(3) & K133 & \nodata & \nodata & \nodata & \nodata & \nodata & \nodata & & 0.32(7) & \nodata \\ 
B59 & 7.3(1) & 0.21(4) & K125 & 7.37(2) & 0.23(2) & 0.23(4) & 11.7(5) & 3.2(4) & 1.2 & & $-0.5(2)$ & $-0.6(1)$ \\ 
B45 & 7.040(4) & 0.144(2) & K124 & \nodata & \nodata & \nodata & \nodata & \nodata & \nodata & & $-0.41(6)$ & \nodata \\ 
S46 & 8.190(4) & 0.204(2) & K123 & 8.32(3) & 0.36(2) & 0.06(4) & 8.1(3) & 4.1(7) & 1.3 & & $-0.87(6)$ & $-1.14(8)$ \\ 
B72 & 8.03(4) & 0.23(4) & \nodata & 7.537(9) & 0.460(7) & $-0.51(2)$ & 19.1(2) & 4.3(1) & 5.3 & & 0.5(1) & 1.0(1) \\ 
S39\tablenotemark{b} & 8.25(3) & 0.43(2) & \nodata & 8.321(7) & 0.398(6) & $-0.29(2)$ & 17.4(1) & 2.2(1) & 1.7 & & 0.50(5) & 0.43(4) \\ 
S37 & 7.96(4) & 0.38(4) & \nodata & \nodata & \nodata & \nodata & \nodata & \nodata & \nodata & & 0.02(6) & \nodata \\ 
B65 & 8.4830(3) & 0.2080(3) & R72 & 8.67(2) & 0.60(3) & $-0.1(1)$ & 10.4(4) & 0.8(2) & 0.9 & & 0.0(2) & $-0.4(3)$ \\ 
S34 & 7.75(4) & 0.50(4) & \nodata & 7.748(9) & 0.451(8) & 0.05(2) & 14.8(1) & 2.0(1) & 1.7 & & 0.26(5) & 0.22(3) \\ 
S32 & 7.70(1) & 0.70(1) & R76 & \nodata & \nodata & \nodata & \nodata & \nodata & \nodata & & $-0.09(1)$ & \nodata \\ 
B67 & 8.590(5) & 0.225(2) & K104 & 8.60(2) & 0.56(3) & 0.18(9) & 13.9(6) & 1.0(2) & 1.4 & & \nodata & \nodata \\ 
B74 & 8.21(2) & 0.37(9) & K101 & 8.07(1) & 0.23(2) & 0.26(6) & 14.2(4) & 2.8(4) & 1.6 & & $-0.56(3)$ & $-0.40(2)$ \\ 
B55 & 7.70(6) & 0.34(6) & \nodata & \nodata & \nodata & \nodata & \nodata & \nodata & \nodata & & 0.0(1) & \nodata \\ 
B54\tablenotemark{c} & 7.57(9) & 0.29(9) & \nodata & 7.66(3) & 0.24(4) & 0.2(2) & 8.4(8) & 2(1) & 0.7 & & $-0.2(2)$ & $-0.4(1)$ \\ 
B87\tablenotemark{b} & 8.150(4) & 0.149(2) & K95 & 8.03(1) & 0.19(2) & 0.26(5) & 10.5(4) & 4.1(5) & 1.2 & & $-1.11(3)$ & $-0.76(5)$ \\ 
B17 & 6.31(3) & 0.10(3) & \nodata & 6.07(2) & 0.15(1) & $-0.03(2)$ & 8.8(5) & 6(1) & 1.0 & & \nodata & \nodata \\
B34 & 7.83(1) & 0.157(5) & K91 & 7.81(3) & 0.20(2) & $-0.08(4)$ & 7.0(4) & 6(2) & 0.9 & & 0.81(8) & 0.9(1) \\ 
B28 & 6.9920(1) & 0.1820(1) & R99 & 7.07(2) & 0.230(5) & 0.2(2) & 9.9(6) & 1.6(7) & 1.7 & & \nodata & \nodata \\ 
B33 & 6.94(4) & 0.34(4) & \nodata & 6.86(2) & 0.38(2) & 0.05(3) & 11.6(3) & 2.1(2) & 1.1 & & \nodata & \nodata \\ 
B41 & 6.5700(1) & 0.1530(1) & R113 & \nodata & \nodata & \nodata & \nodata & \nodata & \nodata & & $-0.1(1)$ & \nodata \\ 
B48 & 6.920(8) & 0.234(3) & K68 & 6.99(1) & 0.24(1) & $-0.03(2)$ & 10.5(3) & 4.4(5) & 1.2 & & \nodata & \nodata \\ 
B115 & 8.27(0) & 0.17(0) & K23 & 8.37(1) & 0.210(8) & 0.03(2) & 11.7(3) & 2.9(3) & 1.5 & & \nodata & \nodata \\
B104 & 9.87(3) & 0.31(1) & K16 & 10.07(2) & 0.51(4) & 0.4(1) & 13.8(7) & 1.4(3) & 1.7 & & 0.31(7) & 0.04(6) \\ 
S1 & 10.240(5) & 0.191(2) & K4 & 10.039(9) & 0.154(6) & $-0.33(1)$ & 9.8(2) & 7.0(8) & 1.5 & & 0.32(3) & 0.77(3)
\enddata
\tablecomments{Negative contraction velocities ({\Vin}~$ <$~0) imply outward motions. Hyperfine structure in {\ammonia} allows for very accurate velocity measurements (see \citealp{Rosolowsky2008} for details).} 
\tablenotetext{a}{Letters of the optically-thin reference ID designating the catalog source -- K for {\NtH} \citep{Kirk2007} and R for {\ammonia} \citep{Rosolowsky2008} surveys -- and the core number in the catalog.}
\tablenotetext{b}{Broad line wings in {\HCO} profiles fit by a Gaussian and removed before modeling with \Hill\ (Section \edit{\ref{subsec:Hill}}).} 
\tablenotetext{c}{Low S/N {\NtD} source (Section \edit{\ref{subsec:lineasymm}}).} 
\label{table:kinematics} 
\end{deluxetable*}

\begin{deluxetable*}{lccccccc}
\tablecolumns{7}
\tablecaption{Summary of Perseus Dense Core Detection, Contraction, Expansion, and Stability Statistics}
\tablehead{
\colhead{} &
\colhead{Detected} &
\colhead{Multiple\tablenotemark{a}} &
\colhead{Single\tablenotemark{a}} & 
\colhead{Contraction\tablenotemark{a}} &
\colhead{Expansion\tablenotemark{a}} & 
\colhead{$M/M_\mathrm{J}$~{\textgreater}~1\tablenotemark{b}} \\
\colhead{} & \colhead{} & \colhead{Gaussian Fits} & \colhead{Gaussian Fit} & \colhead{(\Vin$>0$)} & \colhead{(\Vin$<0$)} & \colhead{}
}
\startdata
Starless       & 31/48  &   16/28  &  12/28  &  5/7 & 2/7  & 6/27  \\
Protostellar  & 41/43  &  22/28   &    6/28  &  8/15 & 7/15  & 16/29 
\enddata
\tablecomments{The percentages of sources well-fit by a single Gaussian and those that show contraction and expansion motions do not sum to 100. Some cores have asymmetric \HCO\, line profiles, but the resulting uncertainty on $V_\mathrm{in}$ from the HILL5 model was larger than the returned $V_\mathrm{in}$ magnitude, and was thus not included in our contraction/expansion analysis.}
\tablenotetext{a}{Numbers are calculated from the number of detected sources in each category, removing the sources with multiple components along the line of sight.}
\tablenotetext{b}{Numbers are calculated from the number of detected sources in each category with continuum data, removing the sources with multiple components along the line of sight. Note that one `static' starless core did not have continuum data. }
\label{table:summary}
\end{deluxetable*}

\section{Core Contraction Analysis}
\label{sec:analysis}

\edit{This section outlines the details of how line asymmetries in {\HCO} profiles are quantified (Section \ref{subsec:lineasymm}) and discusses the use of the {\Hill} model to obtain physical parameters of the dense cores (Section \ref{subsec:Hill}).} To ensure that the observed {\HCO} line profiles are dominated by internal core kinematics, we first identify cores where optically-thin line tracers reveal multiple velocity components that are likely due to secondary sources along the line of sight. We do not detect any secondary cores in the {\NtD} emission profiles, likely because the main hyperfine components are blended together. Through {\NtH} \citep{Kirk2007} and {\ammonia} \citep{Rosolowsky2008} surveys, we identified eight cores with multiple components. Of these, six are protostellar. These sources are specified in Table \ref{table:alltargetsinfo} and were omitted from further analysis. 

There remained seven sources that exhibit line asymmetries in the optically-thick {\HCO} emission, but where examination by eye revealed that the observed systemic velocity was closely aligned with the brightest {\HCO} peak (noted in Table \ref{table:alltargetsinfo}). For a spherically symmetric core undergoing bulk radial motion, the systemic velocity should trace the line profile center \citep{DeVries2005}. It is therefore likely that the asymmetries observed in line profiles with systemic velocities that coincide with the brightest {\HCO} peak are due to multiple cores along the line of sight rather than internal core kinematics. All but one (B46) of these seven cores are found to be protostellar. The secondary sources might still be in their early stages of core evolution such that the dense gas tracers have not yet attained a significant enough abundance to be detectable at our sensitivity or in the above-mentioned {\NtD} and {\ammonia} surveys used in this paper. Another possibility is that the velocity structure seen in {\HCO} is due to lower density gas emitting in filamentary regions surrounding the core, with the dense core moving relative to the surrounding gas. We also remove these sources from further analysis. 

\begin{figure}
\begin{center}
\includegraphics[width=8cm]{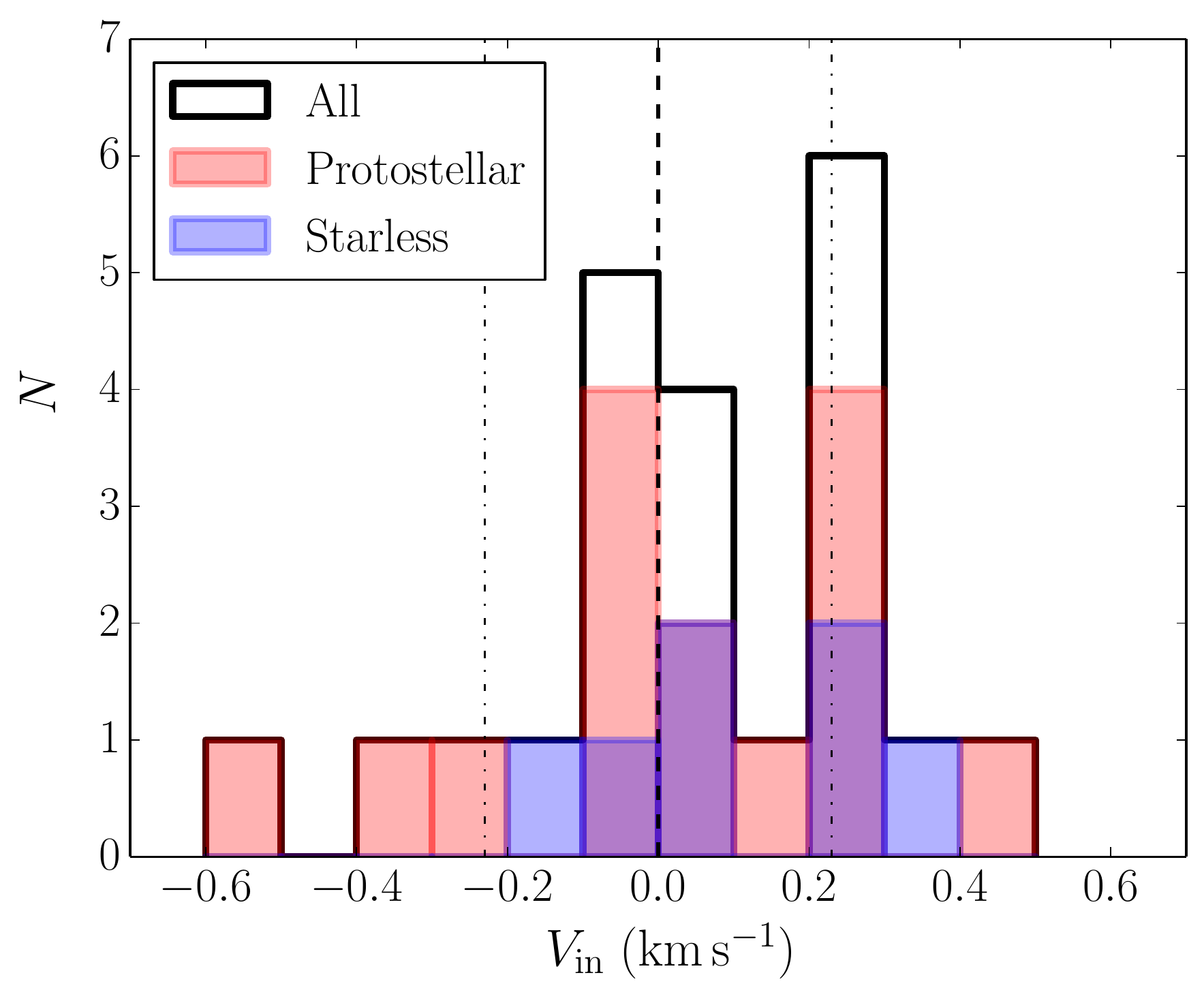}
\caption{Histograms of {\Vin} values found for dense cores modeled with the analytic {\Hill} model. Note that negative contraction velocities ({\Vin}$ < $ 0) imply outward motions. The total distribution is shown in black, with protostellar and starless core distributions shown in red and blue, respectively. Cases where the number of protostars in a bin overlaps the number of starless cores are shown in purple. The vertical dashed line indicates the {\Vin}=0 line, separating expanding sources (left) from collapsing sources (right). The vertical dotted lines indicate the approximate sound speed of the molecular gas (assuming a gas temperature close to a dust temperature $T_d$ = 10 K).}
\label{fig:infallhist}
\end{center}
\end{figure}

\subsection{Line Asymmetry}
\label{subsec:lineasymm}

For a spherically symmetric and static core observed in an optically-thick molecular species, a self-absorption dip appears at the systemic velocity of the core where the gas density and excitation temperature peak, around which the emission line profile is symmetric \citep{DeVries2005}. As the core undergoes collapse, the gas being redshifted becomes self-absorbed by gas with a lower excitation temperature, yielding the classic blue-asymmetric collapse profile \citep{Leung1977, DeVries2005}. If the core is undergoing outward radial motion however, the gas being blueshifted will be locally self-absorbed and the emission line profile will be red asymmetric. 
\edit{The observed line asymmetries are produced by the combination of radial motion and high optical depth and so the kinematic diagnosis is not available with optically-thin lines, for which the shape of the line profile can be ambiguous. Similarly, cores with Gaussian-shaped {\HCO} profiles are assumed to be optically thin and are therefore not included in the core contraction analysis; while they might be contracting, their kinematics cannot be measured without an optically-thick tracer.}

To characterize dense cores that are likely to be collapsing and actively forming stars, we quantify the line asymmetries present in {\HCO} profiles using a dimensionless asymmetry parameter, {\deltaV} (see Equation~\ref{eq:deltaV}). This method is independent of any physical model, is best suited to optically-thick profiles with two well-defined peaks, and is ambiguous for double-peaked profiles with equal intensities \citep{Mardones1997}. Furthermore, \citet{DeVries2005} find that cores with comparatively high velocities begin to flatten out on either the red or blue side of {\HCO} emission for contraction and expansion kinematics, respectively, resulting in a ``shoulder'' profile without two distinct peaks.

To focus on the most asymmetric profiles that are the most likely to be representative of self-absorption caused by radial kinematics, we first quantify temperature asymmetries in the two brightest velocity components in optically-thick {\HCO} emission profiles in \TMB. This is done by measuring the ratio of the blue-red temperature difference of the two main velocity components to the error in that difference, 
\begin{equation} \label{eq:deltaT}
{\delta}_T =  \frac{T_{\mathrm{blue}} - T_{\mathrm{red}}}{\sqrt{     {\sigma}_{T_{\mathrm{blue}}}^2 + {\sigma}_{T_{\mathrm{red}}}^2  }} \, ,
\end{equation}
using the Gaussians line profile fit results discussed in Section \edit{\ref{sec:results}}. This parameter has been used on its own to trace contraction \citep[e.g.,][]{Myers1995, Empr2009}. The parameters $T_\mathrm{blue}$ and $T_\mathrm{red}$ are the main-beam temperatures of the blue and red velocity components, while ${\sigma}_{T_\mathrm{blue}}$ and ${\sigma}_{T_\mathrm{red}}$ are their uncertainties, respectively. We use the uncertainty from our Gaussian fitting procedure as the uncertainty measurement rather than the rms noise \citep{Mardones1997} and restrict our analysis to those sources characterized by $|${\deltaT}$|$~{\textgreater}~2. 

\edit{Considering only sources with no known or suspected secondary core along the line of sight (hereafter single cores)}, we find 20 cores with significant temperature asymmetries in their {\HCO} line profiles. 
The sources where $|${\deltaT}$|$~{\textgreater}~2 have \deltaV\ entries in Column 11 of Table \ref{table:kinematics}, with the running source name from Table~\ref{table:alltargetsinfo} indicated in Column 1. 
\edit{As discussed in Section \ref{sec:results}, where cores were not detected in our optically-thin line tracer {\NtD}, we used optically-thin measurements of the core's local standard-of-rest velocity and line width from, in order of precedence, {\NtH} \citep{Kirk2007} and {\ammonia} \citep{Rosolowsky2008} to calculate \deltaV.} 
Note that Table \ref{table:kinematics} also includes a different subset of sources where the line profiles were fit by an analytic collapse model, described below \edit{in Section \ref{subsec:Hill}}.

A small fraction of \HCO-detected \edit{single} cores (20/56, or 36\,\%) show significant line asymmetries in their line profiles indicative of radial motions. A higher fraction of \edit{single} cores with strong temperature asymmetries in their {\HCO} profiles was found to be protostellar (12/20, or 60\,\%) compared to their starless counterparts (8 of 20, or 40\,\%) as well as the protostellar cores that do not exhibit line asymmetries (16/36, or 44\,\%). Given that the \HCO-detected \edit{single} cores are equally divided into protostellar (28 of 56) and starless (28 of 56) cores, we conclude that protostellar cores are more likely to exhibit line asymmetries than their starless counterparts.

\subsection{Analytic HILL5 Modeling}
\label{subsec:Hill}

\begin{figure*} 
\begin{center}
\includegraphics[width=8.5cm]{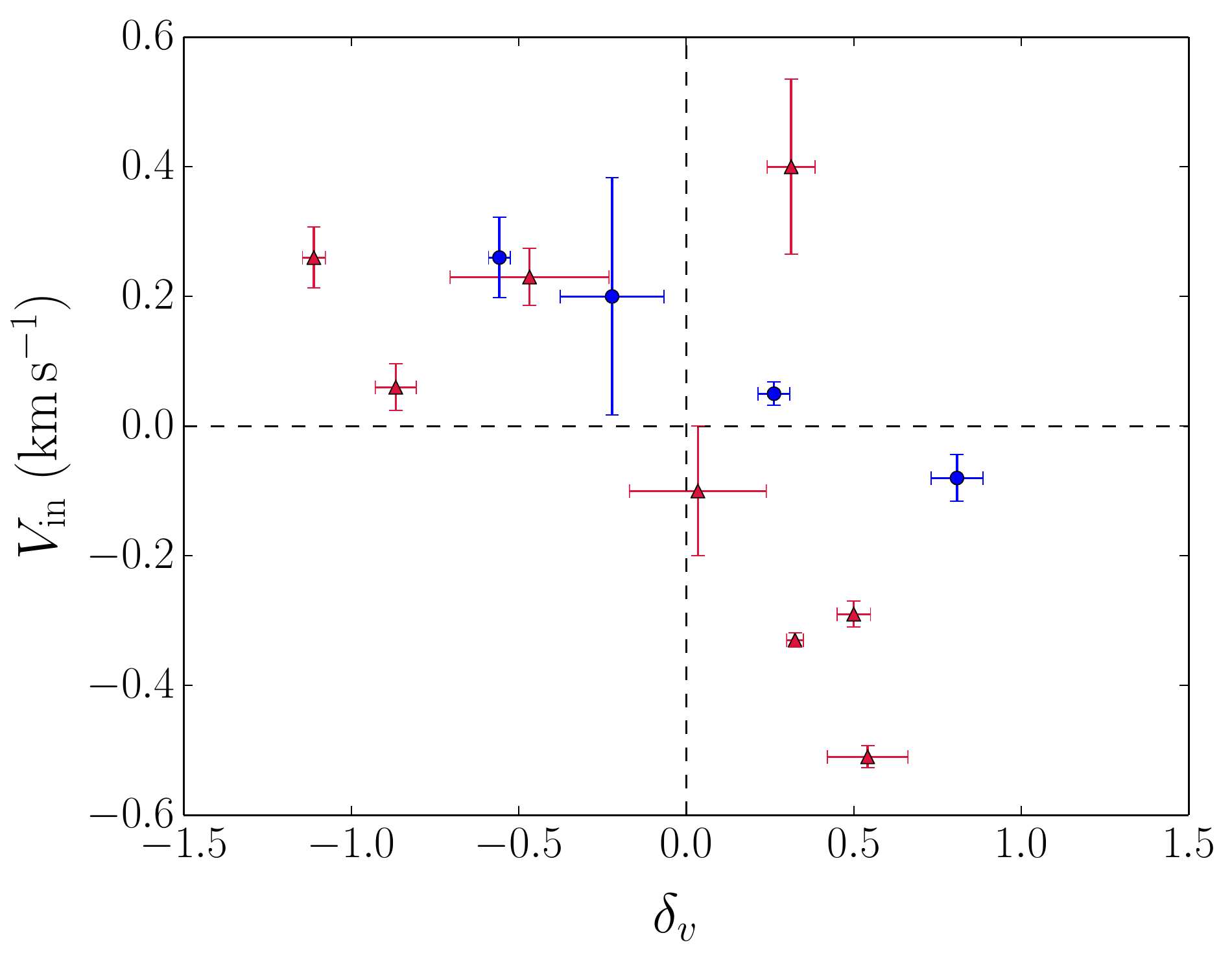}
\hspace{0.5cm}
\includegraphics[width=8.5cm]{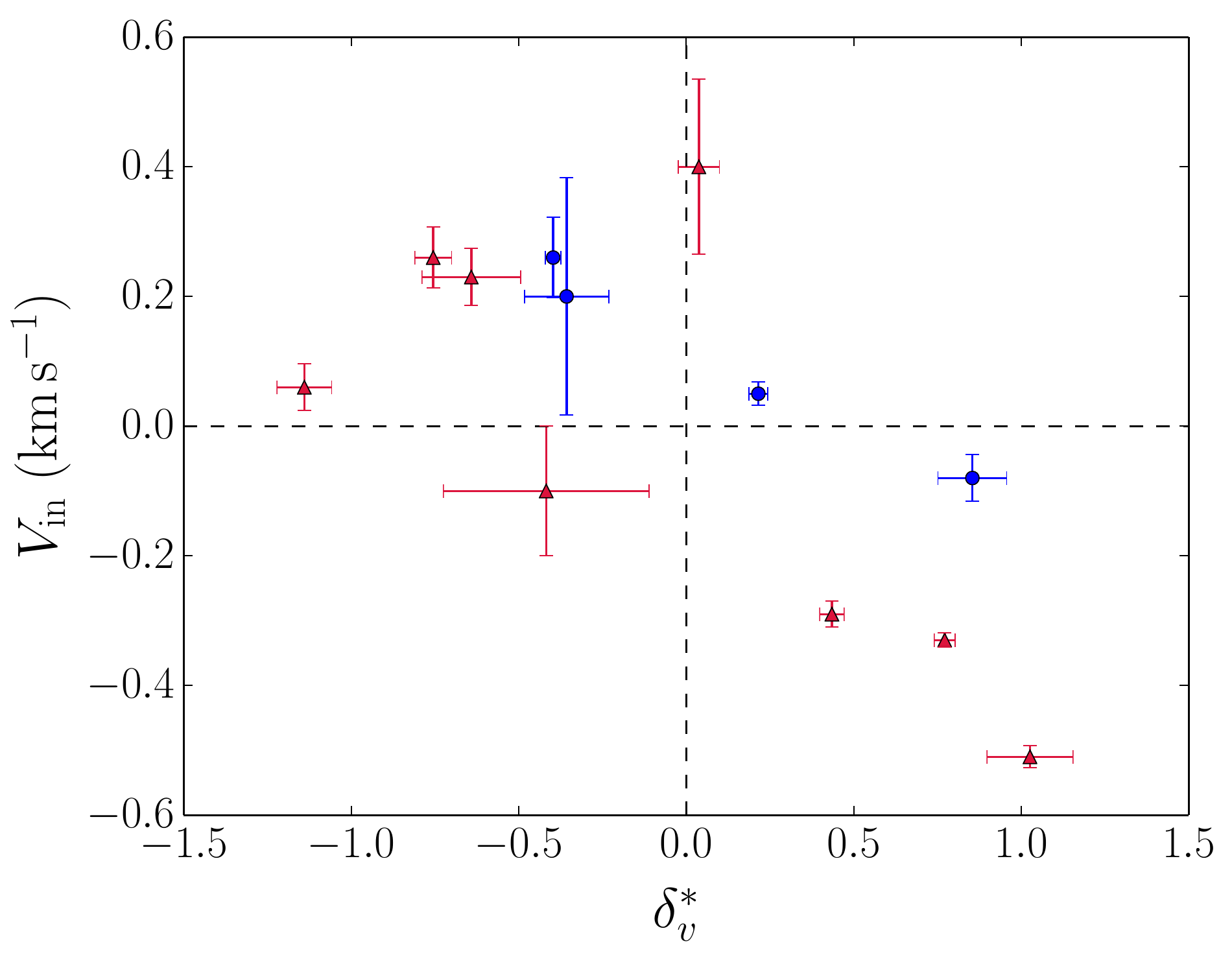}
\caption{Left: Modeled contraction velocities for asymmetric {\HCO} profiles versus the calculated asymmetry parameter, {\deltaV}. The Pearson correlation coefficient between these two parameters is {\rhoPCC}$=-0.59$. Note that negative contraction velocities ({\Vin}$ < $ 0) imply outward motions. In both plots, starless cores are indicated with (blue) circles and protostellar cores with (red) triangles.
Right: Modeled contraction velocities for asymmetric {\HCO} profiles versus the modified calculated asymmetry parameter, {\deltaVstar}, which uses the model-derived systemic velocity in replacement of that observed with Nitrogen-based optically-thin tracers. The Pearson correlation coefficient between these two parameters is {\rhoPCC}$=-0.69$. 
\notetoeditor{These figures should remain side-by-side in print.}}
\label{fig:infallcomparison}
\end{center}
\end{figure*}

\begin{figure}
\begin{center}
\includegraphics[width=8.5cm]{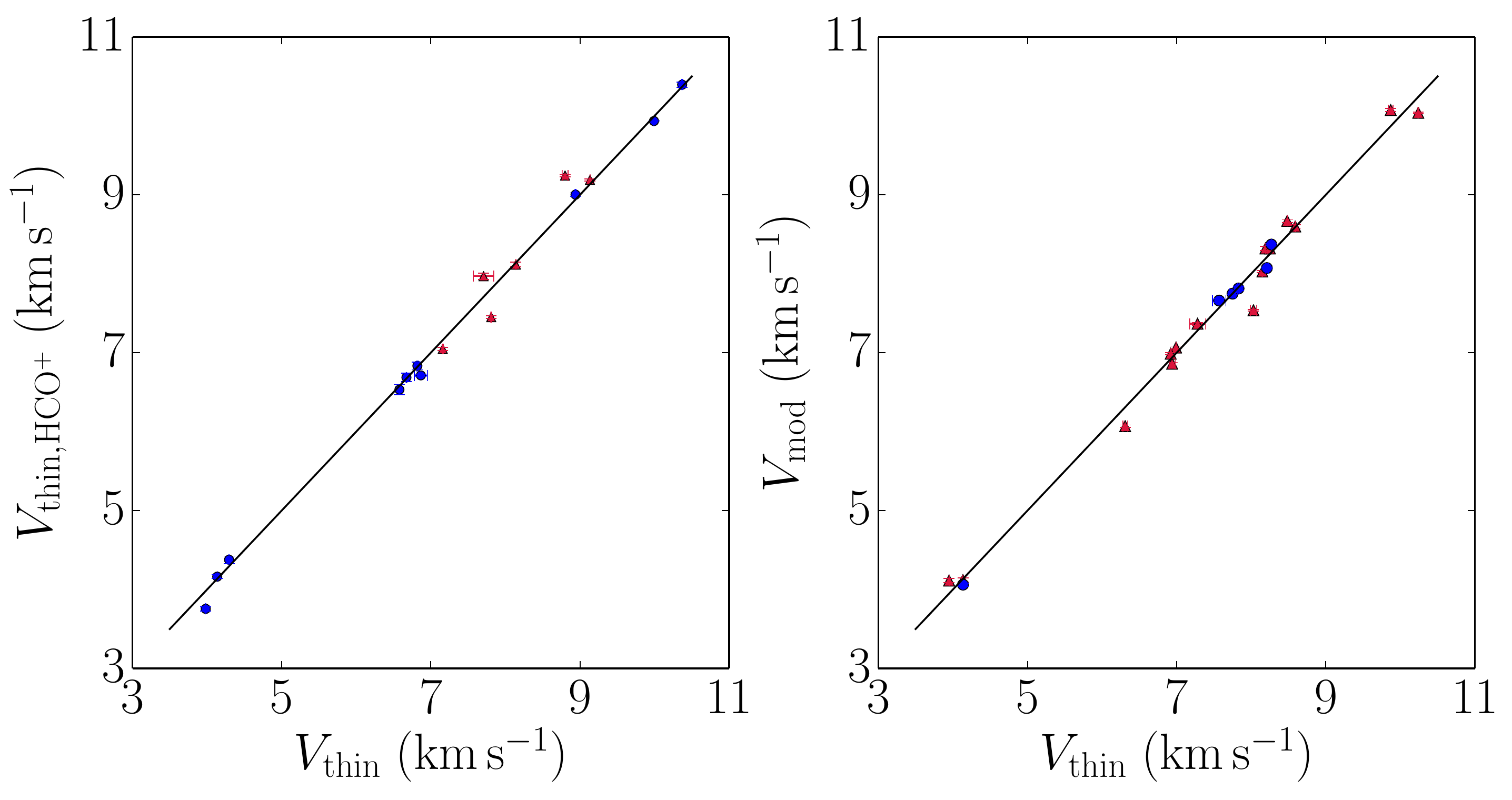}
\caption{Correlations between the systemic velocity (\Vcent) of optically-thin {\HCO} profiles (left) and model-derived systemic velocities (\Vmod) of optically-thick {\HCO} profiles (right), with the systemic velocity obtained from their optically-thin counterparts (\Vthin). The Pearson correlation coefficient between the two measures is {\rhoPCC}=1.00 for both {\Vcent} (top) and {\Vmod} (bottom). Starless cores are indicated with (blue) circles and protostellar cores with (red) triangles. The solid diagonal lines represents the 1:1 correspondence.}
\label{fig:movelvsthinsystemic}
\end{center}
\end{figure}

\begin{figure}
\begin{center}
\includegraphics[width=7cm]{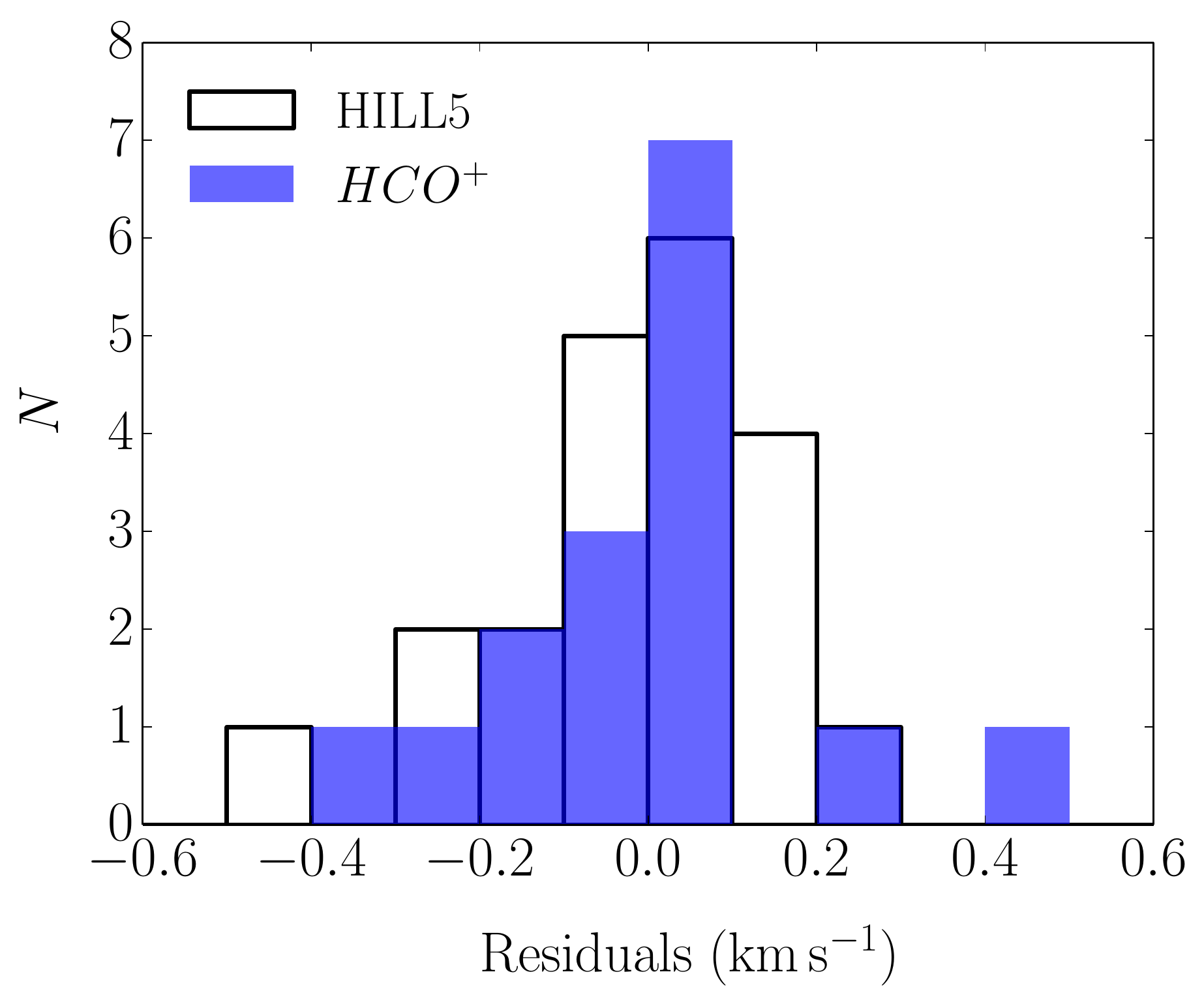}
\caption{Distributions of residuals of the centroid velocities for the optically-thin {\HCO} profiles (blue) and the modeled systemic velocity for asymmetric {\HCO} profiles (white) with respect to the systemic velocity provided by the Nitrogen-based optically-thin molecular tracers.}
\label{fig:residualhist}
\end{center}
\end{figure}

Since {\deltaV} does not provide radial velocities of the cores, we fit the observed {\HCO} line profiles with a one-dimensional analytic model of a collapsing core \citep{DeVries2005}. We use the {\Hill} model, which assumes that the excitation temperature follows a linear relationship with optical depth along the line of sight, increasing toward the core centers. The {\Hill} model assumes two layers of gas are moving toward or away from each other at a constant velocity within the core, and best reproduces the physical parameters of the simulated collapsing cores that exhibit double-peaked, asymmetric line profiles.

We use the {\sc idl} {\tt MPFIT} chi-squared minimization routine \citep{Markwardt09} to fit the {\Hill} model with five free parameters: the line-of-sight systemic velocity of the core with respect to the local standard of rest (\Vmod), the velocity dispersion of the molecular species ($\sigma$), the contraction velocity ({\Vin}), the peak excitation temperature of the core ({\Tex}), and the line center optical depth ($\tau$). The {\Hill} model was run on all \edit{\HCO-detected single cores}. We considered a significant result to be one for which the magnitude of the contraction velocity was larger than its uncertainty. Cores successfully fit by the {\Hill} model have entries in Columns 5 to 10 in Table~\ref{table:kinematics}.

The analytic model is defined for \emph{isolated, starless} cores and does not account for features that may be present in protostellar cores such as internal
\edit{heating, molecular outflows, or dense gas in a filamentary or clustered environment causing a secondary velocity component in the line profile. For protostellar cores in which there are both core contraction and an outflow, for example, the {\Hill} model of simple, spherical collapse would not be an ideal model, and the ability to detect a contraction (expansion) signature in the line profile and then obtain an accurate measure of the contraction (expansion) velocity would depend on the orientation of the core and its outflow relative to our line-of-sight. While the {\Hill} model might therefore not be the best model for our data set as a whole, more detailed models for protostellar collapse including secondary components are nevertheless beyond the scope of this paper.}

To reduce systematic errors with cores that exhibit features likely to arise from YSOs, such as high velocity line wings, we fit broad Gaussians to the emission spectra of two sources to remove these broad features before proceeding with the analytic modelling (noted in Table \ref{table:kinematics}). Note that negative contraction velocities ($V_\mathrm{in} < 0$) imply outward motions. Since our observations are single pointings, we cannot distinguish between cores with negative contraction velocities that are being powered by molecular outflow jets and those that are expanding or oscillating about an equilibrium \citep{lada03,lee11,broderick10}. We argue that protostellar cores with negative contraction velocities, especially those with signatures of high velocity line wings in their emission profiles, are likely to be powered by YSO-driven molecular outflows. Indeed, both cores from which broad line wings were removed were found to be protostellar. 

We show an example spectrum where an expansion profile for source S39 was modeled using the {\Hill} model in Figure \ref{fig:examplespectra} (top). This source was one of the two (protostellar cores) whose broad line wings were removed, and it can be seen that the {\Hill} analytic model does a very good job at fitting the line profile for this source. The analytic model determines {\Vin}~{\textless}~0, as one might expect from the strong red asymmetry seen in {\HCO}. The solid and dashed vertical lines indicate the systemic velocity traced by {\NtD} and that provided by the analytic model, respectively; these agree very closely. 

The physical parameters provided by the {\Hill} model for 22 cores are listed in Columns 5 through 10 of Table \ref{table:kinematics}. The model-derived systemic velocity ({\Vmod}) is listed in Column 5, the velocity dispersion ($\sigma$) in Column 6, the contraction velocity ({\Vin}) in Column 7, the excitation temperature ($T_\mathrm{ex}$) in Column 8, the optical depth ($\tau$) in Column 9, and the reduced chi-squared goodness of fit parameter ({\redchisq}) in Column 10. Note that negative contraction velocities ({\Vin}~{\textless}~0) imply outward motions. The resulting analytic model fits are shown in Figure \ref{fig:modelplots} in the Appendix.

Of the 22 cores that reveal significant radial motion when modeled with \Hill, and hence have results listed in Table~\ref{table:kinematics}, 13 have {\Vin}~{\textgreater}~0 indicative of contracting motions while the remaining 9 show expanding motions. We find that protostellar sources comprise the majority (15/22, or 68\,\%) of cores with measured radial velocities. We find that protostellar cores are almost twice as likely to show contraction motions (8/13 versus 5/13 for starless cores), while protostellar cores are significantly more likely to show expanding motions (7/9 versus 2/9 for starless cores), likely due to expansions driven by embedded YSOs. Figure \ref{fig:infallhist} shows the distribution of {\Vin} for protostellar and starless cores, in red and blue, respectively, with the entire distribution shown in black. We discuss the range in contraction speeds further in the next section, but note that on average, protostellar sources are found to have both higher contraction and expansion speeds on average than the starless cores. The statistics of the model-derived contraction velocity results are summarized in Table \ref{table:summary}.

\section{Discussion}
\label{sec:discussion}

\subsection{Comparison of {\deltaV} and \Vin}
\label{subsec:comparison}

To test the efficacy of using the asymmetry parameter of \citet{Mardones1997} as an indicator of dense cores undergoing collapse in Perseus, we compare directly {\deltaV} and {\Vin} in Figure \ref{fig:infallcomparison} (left). As one might hope, the line asymmetries quantified by {\deltaV} approximately anti-correlate with the contraction velocity {\Vin} derived from the analytic model. However, there is a broad scatter between these two contraction tracers and there are two sources in the upper-right quadrant where {\deltaV} and {\Vin} disagree on the direction of radial motion. 

In some cases, the multiple Gaussian fitting routine does not accurately reproduce the velocity of the brightest velocity component in the optically-thick emission profile. 
\edit{This can be most problematic for profiles that have narrow features or where the velocity components are not well separated such as can happen for very broad profiles or profiles with shoulders.}
This results in {\Vthick} being redshifted or blueshifted from the velocity of the brightest component and, in some cases, the opposite sign of {\deltaV} than expected. Another source of variation between {\deltaV} and {\Vin} occurs when the observed systemic velocity does not perfectly coincide with the line profile center, which can more significantly affect {\deltaV} in cases where the emission line profile shows narrow features or shoulders as mentioned above. We note that without removing sources with known or suspected multiple cores along the line of sight, the correlation is substantially worse, emphasizing the necessity of using the systemic velocity traced by optically-thin tracers as an additional constraint on contraction analyses. 

The self-absorption dip in optically-thick emission occurs at the systemic velocity of the core, which should trace the line profile center if the line asymmetry is due to radial motion. To investigate the effect that the subtle systemic velocity offsets from the {\HCO} line profile centers have on the correlation between {\deltaV} and {\Vin}, we repeat the correlation analysis using a modified asymmetry parameter, {\deltaVstar}. This modified asymmetry parameter replaces the systemic velocity {\Vthin} with the {\Hill} model-derived systemic velocity {\Vmod} in Equation~\ref{eq:deltaV}. Figure \ref{fig:infallcomparison} (right) shows {\Vin} versus {\deltaVstar}. The anti-correlation is slightly stronger; {\deltaV} identifies core contraction relatively well when the optically-thin systemic velocity accurately traces the line profile center of the optically-thick emission profile (or vice versa), as is explored further in the following section. 
\edit{We note that two sources remain in the lower-left or upper-right quadrants, within uncertainties (B65 and S34, respectively). These sources have broad profiles that are not Gaussian, but do not have an obvious self-absorption dip clearly demarcating the blue and red velocity components. While they pass our {\deltaT} test, in these cases the measured {\Vthick} might not accurately reproduce the velocity at the profile peak, resulting in {\deltaVstar} and {\Vin} disagreeing in the direction of radial motion.}

\edit{As discussed in Section \ref{subsec:lineasymm}, cores with {\HCO} velocity profiles that are well approximated by a single Gaussian are assumed to be optically thin and are not included in the core contraction analysis. These {\HCO} profiles are used instead to investigate whether there is any systematic offset between the apparent {\HCO} velocity ({\Vcent}) and the velocity {\Vthin} from our Nitrogen-based optically-thin tracers that might skew our {\deltaV} analysis.} 
Figure \ref{fig:movelvsthinsystemic} (left) shows {\Vcent} against {\Vthin}, indicating good agreement.
In Figure \ref{fig:movelvsthinsystemic} (right) we show a similar correlation between {\Vmod}, the systemic velocity provided by {\Hill}, and {\Vthin}. 
Figure \ref{fig:residualhist} shows the distribution of residuals from these correlations. The median residual is 0.013\,\kms\ and 0.009\,\kms\ for {\Vcent} and {\Vmod}, respectively, while the distributions of residuals have standard deviations of 0.17\,\kms\ and 0.16\,\kms, respectively. Although the line profile centers are well measured by the Nitrogen-bearing molecular tracers in both the optically-thin and optically-thick {\HCO} samples, the dispersions of the distributions of residuals are not insignificant compared to measured contraction velocities and asymmetry parameters.

\subsection{Contraction and Core Stability}
\label{subsec:stability}

The stability of dense cores requires the counterbalancing of their own self-gravity and any external pressures by their internal means of support. Here we investigate the ratio of the core masses to their Jeans mass (\MJ), the critical mass that a gas cloud can support (not collapsing under its own self-gravity) assuming only thermal support is important. Higher ratios of core mass to Jeans mass (\MMJ) should thus be indicative of cores more likely to exhibit ongoing star formation where $M/M_J=1$ defines the critical self-gravitating limit. The Jeans mass is expected to be particularly relevant in cold, quiescent dense cores where the turbulent motions from the surrounding molecular cloud have dissipated substantially. In agreement with this, in a survey of prestellar cores in the Ophiuchus molecular cloud, \citet{simpson11} show that {\MMJ} is a good predictor of core instability, finding that between 40\,\% to 60\,\% of cores with $M/M_\mathrm{J} > 1$ show evidence of line asymmetries that they attribute to contraction. Similarly, in a sample of highly evolved starless cores across multiple molecular clouds, \citet{Schnee2013} show that contraction motions are more likely in cores with $M/M_\mathrm{J} > 1$.

We first determine masses for all cores detected in {\HCO}, with one exception (D38), using 1.1\,mm \citep{Enoch2006} and 850\,\micron~\citep{Kirk2006} surveys of Perseus (note that we use the fluxes reported in \citealt{Kirk07e}, and that the continuum emission in these sources does not necessarily match the extent of the molecular emission). For each source, we identify the nearest (sub)mm core within 15\arcsec\ (the approximate beam size) of our targets and calculate the core mass using
\begin{equation}
\label{eq:mass_estimate}
M = \frac{d^2 F_\nu}{\kappa_\nu B_\nu (T_d)} \, ,
\end{equation}
where $d$ is the distance to Perseus, $F_\nu$ is the total observed (sub)mm flux density, $\kappa_\nu$ is the dust opacity at frequency $\nu$, and $B_\nu (T_d)$ is the Planck function at dust temperature $T_d$. We use a dust opacity $\kappa = 0.0114$\,cm$^2$\,g$^{-1}$ at 1.1\,mm (as did \citealp{Enoch2006}) 
and $\kappa = 0.018$\,cm$^2$\,g$^{-1}$ at 850\,{\micron} in accordance with the \citet{ossen94} dust models with thin ice mantles and a gas density of 10$^6$\,cm$^{-3}$. These two values are consistent with one another for an emissivity spectral index of $\sim 2$, typical for dense cores \citep{Schnee2010}. \citet{Rosolowsky2008} determine the gas kinetic temperature, \tk\,, for many of the Perseus cores. 
Where our target positions agree within 15\arcsec\ we set $T_d = T_\mathrm{K}$,
and where there is no overlap we set $T_d = 11$~K, the mean \tk\ for all detected cores in Perseus. At the densities typical of most of the cores observed here \citep[$n \gtrsim 10^{4.5}$\,cm$^{-3}$;][]{Kirk2006,Enoch2006} we expect the gas and dust to be coupled \citep{doty97,goldsmith01}. This assumption is therefore not a significant source of uncertainty in the final masses. The single starless core with no reported continuum flux density, and hence no calculated core mass, showed no line asymmetry in the detected {\HCO} line. 

\begin{deluxetable}{lcccccccc}
\tablecolumns{9}
\tablecaption{Continuum Masses and Stability of the Perseus Cores}
\tablehead{
\colhead{} & \multicolumn{2}{c}{Continuum} & \colhead{} & \multicolumn{2}{c}{Gas} & \colhead{} & \multicolumn{2}{c}{Jeans Analysis} \\
\cline{2-3} \cline{5-6} \cline{8-9} \\ 
\colhead{Source} &
\colhead{$R_\mathrm{eff}$} &
\colhead{$M$} & \colhead{} &
\colhead{$T_\mathrm{K}$\tablenotemark{a}} &
\colhead{${\sigma}_\mathrm{NT}$} & \colhead{} &
\colhead{$M_\mathrm{J}$} &
\colhead{$M/M_\mathrm{J}$} \\
\colhead{} &
\colhead{(pc)} &
\colhead{(\Msun)} & \colhead{} &
\colhead{(K)} &
\colhead{(\kms)} & \colhead{} &
\colhead{(\Msun)} & 
\colhead{} \\
\colhead{(1)} &
\colhead{(2)} &
\colhead{(3)} & \colhead{} &
\colhead{(4)} &
\colhead{(5)} & \colhead{} &
\colhead{(6)} &
\colhead{(7)} \T
}
\startdata
\multicolumn{9}{c}{starless} \B\\
\cline{1-9}
 B16 &  0.06 &  0.3 & &  9.2 &  0.09 & &  1.4 &  0.2 \T\\
 B31 &  0.11 &  0.9 & & 10.0 &  0.12 & &  3.1 &  0.3 \\
 B11 &  0.13 &  2.8 & & 10.3 &  0.11 & &  3.7 &  0.7 \\
 B27 &  0.10 &  1.7 & & 11.2 &  0.18 & &  2.9 &  0.6 \\
 B24 &  0.04 &  4.5 & &  9.1 &  0.13 & &  1.1 &  4.0 \\
 B23 &  0.06 &  1.2 & &  9.0 &  0.14 & &  1.5 &  0.8 \\
 B20 &  0.09 &  0.7 & & 11.0 &  \nodata & &  2.7 &  0.3 \\
 S50 &  0.03 &  1.3 & & 10.5 &  0.17 & &  0.8 &  1.6 \\
 B43 &  0.09 &  1.1 & & 10.6 &  0.13 & &  2.6 &  0.4 \\
 B71 &  0.07 &  1.0 & & 13.6 &  0.23 & &  2.7 &  0.4 \\
 S37 &  0.04 &  2.2 & & 14.8 &  0.35 & &  1.5 &  1.5 \\
 S34 &  0.04 &  3.8 & & 11.0 &  \nodata & &  1.2 &  3.1 \\
 B74 &  0.03 &  0.4 & & 15.5 &  0.41 & &  1.1 &  0.4 \\
 B55 &  0.06 &  1.2 & & 11.0 &  \nodata & &  1.8 &  0.7 \\
 B54 &  0.06 &  0.7 & & 11.0 &  \nodata & &  1.9 &  0.4 \\
 B34 &  0.10 &  1.5 & & 10.5 &  0.13 & &  3.0 &  0.5 \\
 B37 &  0.12 &  4.5 & & 10.2 &  0.11 & &  3.4 &  1.3 \\
 B38 &  0.11 &  2.2 & & 11.2 &  0.22 & &  3.3 &  0.7 \\
 B40 &  0.09 &  2.2 & & 10.1 &  0.12 & &  2.3 &  0.9 \\
 B41 &  0.07 &  0.9 & & 10.0 &  0.14 & &  1.8 &  0.5 \\
 B42 &  0.07 &  0.9 & &  9.9 &  0.18 & &  1.8 &  0.5 \\
 B44 &  0.07 &  0.9 & &  9.7 &  0.10 & &  1.8 &  0.5 \\
 B77 &  0.08 &  0.8 & & 11.0 &  \nodata & &  2.3 &  0.3 \\
B115 &  0.03 &  1.1 & & 12.9 &  0.16 & &  1.2 &  0.9 \\
B103 &  0.07 &  0.8 & & 10.8 &  0.10 & &  2.0 &  0.4 \\
B116 &  0.13 &  2.7 & & 10.7 &  0.08 & &  3.7 &  0.7 \\
  S2 &  0.03 &  1.4 & & 10.1 &  0.17 & &  0.9 &  1.6 \B\\
\cline{2-9}
mean:  & 0.07 &  1.6 & & 10.9 & 0.13 & &  2.1 & 0.9 \T\B\\
\cline{1-9}
\multicolumn{9}{c}{protostellar} \T\B\\
\cline{1-9}
 B15 &  0.05 &  0.4 & &  9.2 &  0.14 & &  1.3 &  0.3 \T\\
 B30 &  0.04 &  3.1 & & 12.1 &  0.15 & &  1.3 &  2.4 \\
 S57 &  0.04 &  3.1 & & 11.2 &  0.14 & &  1.4 &  2.3 \\
  B5 &  0.06 &  0.6 & & 13.1 &  0.26 & &  2.1 &  0.3 \\
  B2 &  0.04 &  1.8 & & 11.4 &  0.15 & &  1.1 &  1.6 \\
  B3 &  0.04 &  1.6 & & 11.9 &  0.20 & &  1.2 &  1.3 \\
 B59 &  0.03 &  1.5 & & 12.4 &  0.20 & &  1.1 &  1.3 \\
 B45 &  0.06 &  0.6 & & 10.5 &  0.13 & &  1.9 &  0.3 \\
 S46 &  0.04 &  2.2 & & 11.7 &  0.17 & &  1.3 &  1.7 \\
 S44 &  0.04 &  1.9 & & 11.7 &  0.18 & &  1.2 &  1.7 \\
 B64 &  0.06 &  0.4 & & 11.3 &  0.10 & &  1.7 &  0.2 \\
 B56 &  0.01 &  0.2 & & 12.5 &  0.19 & &  0.5 &  0.4 \\
 B72 &  0.05 &  4.8 & & 16.4 &  0.25 & &  2.4 &  2.0 \\
 S39 &  0.04 &  5.8 & & 16.3 &  0.28 & &  1.9 &  3.0 \\
 S35 &  0.04 &  3.3 & & 11.0 &  \nodata & &  1.3 &  2.5 \\
 B65 &  0.08 &  2.2 & & 12.6 &  0.19 & &  2.6 &  0.8 \\
 S32 &  0.04 &  3.0 & & 11.0 &  \nodata & &  1.2 &  2.5 \\
 B67 &  0.05 &  4.0 & & 14.3 &  0.21 & &  1.9 &  2.1 \\
 B87 &  0.02 &  0.9 & & 10.5 &  0.10 & &  0.7 &  1.2 \\
 B17 &  0.09 &  1.9 & & 10.0 &  0.13 & &  2.4 &  0.8 \\
 B28 &  0.05 &  1.4 & & 11.4 &  0.17 & &  1.7 &  0.9 \\
 B33 &  0.04 &  7.2 & & 11.7 &  0.23 & &  1.2 &  5.8 \\
 B48 &  0.08 &  0.8 & & 11.9 &  0.24 & &  2.5 &  0.3 \\
B112 &  0.04 &  2.5 & & 11.7 &  0.18 & &  1.4 &  1.8 \\
B106 &  0.06 &  0.8 & & 13.0 &  0.26 & &  2.3 &  0.3 \\
B102 &  0.08 &  0.8 & & 11.0 &  0.21 & &  2.3 &  0.3 \\
B100 &  0.11 &  1.6 & & 10.5 &  0.14 & &  3.2 &  0.5 \\
B104 &  0.05 &  0.6 & & 16.0 &  0.33 & &  2.1 &  0.3 \\
  S1 &  0.03 &  1.0 & & 11.7 &  0.15 & &  0.9 &  1.2 \B\\
\cline{2-9}
mean: & 0.05 &  2.1 & & 12.3 & 0.19 & &  1.6 & 1.4 \T\\
\enddata
\tablenotetext{a}{For cores with no overlap with \citet{Rosolowsky2008}, we \\set $T_d = 11$~K, the mean \tk\, for all detected cores in Perseus.}
\label{table:mm} 
\end{deluxetable}

We next determine the Jeans mass for each core following \citet{Sadavoy2010}:
\begin{equation} 
\label{eq:JeansMass}
M_\mathrm{J} = 1.9\left(\frac{T_d}{10~\mathrm{K}}\right)\left(\frac{R_J}{0.07~\mathrm{pc}}\right){\mathrm{M}_{\odot}} \, .
\end{equation}
Here, we set $R_\mathrm{J} = R_\mathrm{eff}$ from the continuum surveys to estimate the Jeans radius, where $R_\mathrm{eff}$ is the geometric mean of the major and minor FWHM axes of the continuum clumps. In Table~\ref{table:mm} we list the running source name in Column 1, the effective core radius ($R_\mathrm{eff}$) from the continuum surveys in Column 2, the derived core mass ($M$) in Column 3, the gas kinetic temperature (\tk\,) and non-thermal velocity dispersion (${\sigma}_\mathrm{NT}$) where measured in Columns 4 and 5, respectively, and the Jeans mass (\MJ) and ratio of core mass to Jeans mass (\MMJ) in Columns 6 and 7, respectively.

\begin{figure*}
\begin{center}
\includegraphics[width=8cm]{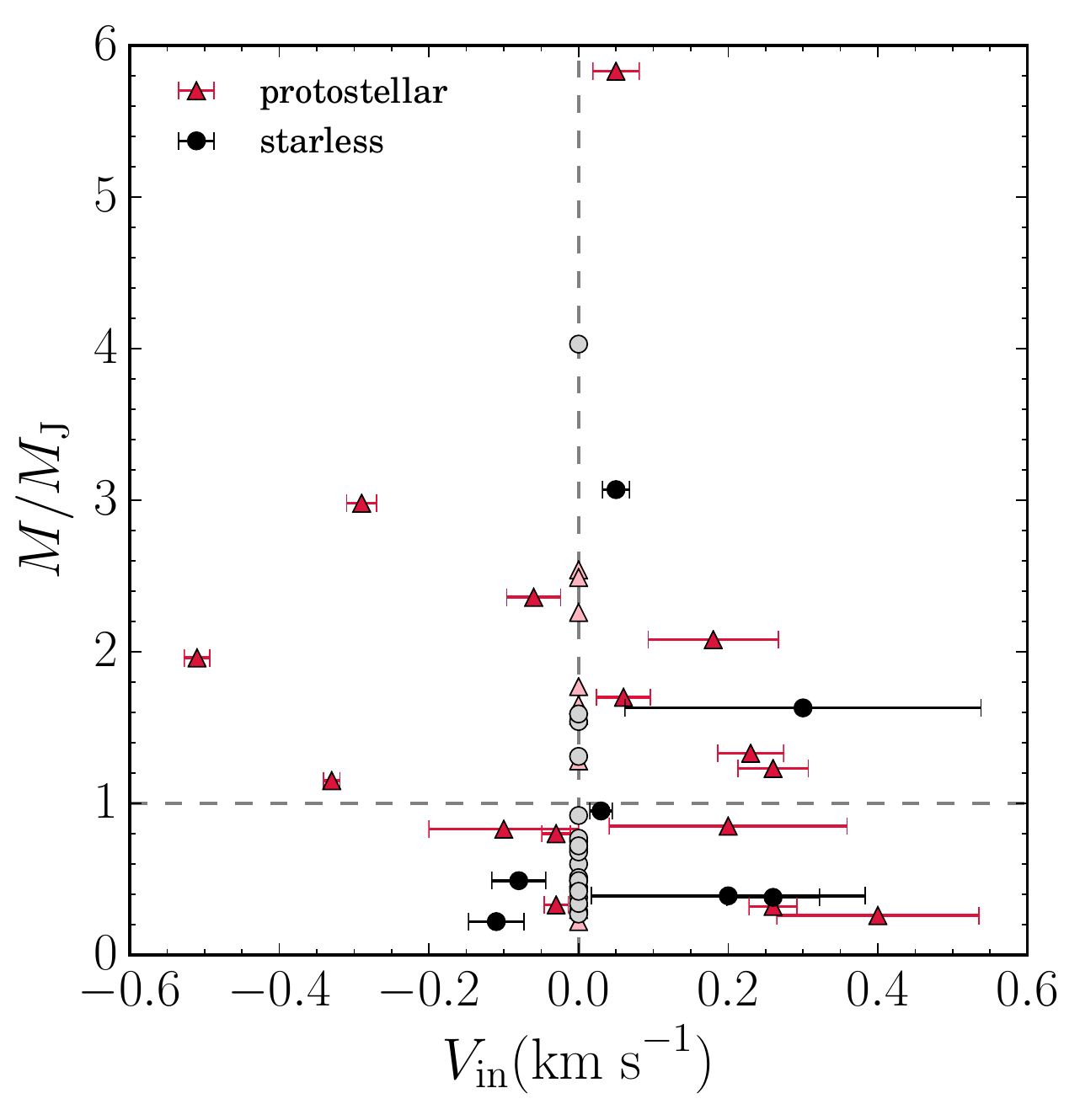} \includegraphics[width=8.2cm]{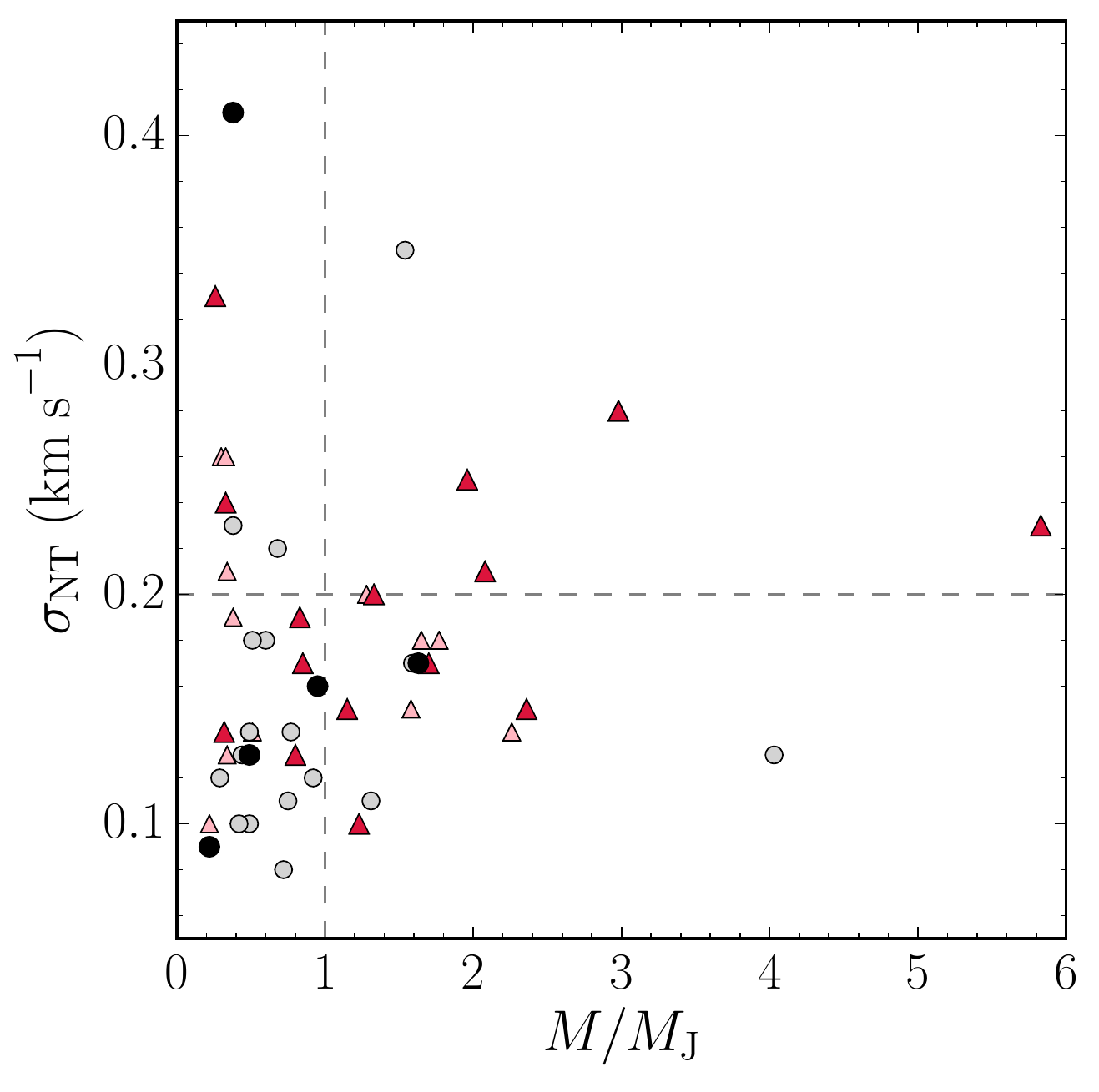}
\caption{Left: Scatterplot of the core mass as a fraction of the Jeans mass versus the modeled contraction velocity ({\Vin}). Note that negative contraction velocities ({\Vin}$ < $ 0) imply outward motions. Cores with no detected contraction or expansion motions have been assigned {\Vin} = 0. 
Right: Scatterplot of the non-thermal line width as measured by optically-thin {\ammonia} emission lines versus the core mass as a fraction of Jeans mass.
Horizontal line indicates the sound speed for H$_2$ at 11~K.
In both plots, circles indicate starless cores; black circles indicate starless cores with measured radial (contraction or expansion) motions and grey circles indicate ``static" starless cores. 
Triangles indicate protostellar cores; red triangles indicate protostellar cores with measured radial (contraction or expansion) motions and light red triangles indicate ``static'' protostellar cores.}
\label{fig:mmcorrelations}
\end{center}
\end{figure*}

The protostellar cores have, on average, larger masses and smaller effective radii than the starless cores. Our starless sample has an average core mass of 1.6~{\Msun} while our protostellar sample has an average core mass of 2.1~{\Msun}, with average $R_\mathrm{eff}$ of 0.07\,pc and 0.05\,pc, respectively. Over all the Perseus cores with \HCO\ detections, we find an average $M_\mathrm{J} = 1.9$\,\Msun. The starless core population has a greater average Jeans mass than the protostellar cores (2.1 {\Msun} versus 1.6 {\Msun}, respectively). The protostellar sample of cores has, on average, higher {\MMJ} values than do their stares counterparts (1.4 and 0.9, respectively). Approximately half of the protostellar cores have $M/M_\mathrm{J} > 1$ (17 of 30), compared with only 19\ \% (5 of 26) of starless cores (see Table \ref{table:summary}; note that one starless core does not have an associated core mass). These five supercritical starless cores are interesting candidates for future studies on evolving dense cores. 

Figure \ref{fig:mmcorrelations} (left) shows {\MMJ} as a function of {\Vin} for the Perseus cores. The horizontal dashed line at $M/M_\mathrm{J} = 1$ shows the transition between sub-critical and super-critical thermally-supported cores. In the figure protostellar and starless cores are identified separately (red triangles and black circles, respectively). Cores that were found not to show contracting or expanding motions above our threshold in the {\Hill} model fitting are assigned a radial velocity of zero, with no error bars, and are also separated into protostellar and starless populations. The figure shows that we find inward and outward motions over the entire range of core {\MMJ} ratios, and in addition many cores with $M/M_\mathrm{J} > 1$ without substantial inward or outward motions. Approximately the same fraction of cores with $M/M_\mathrm{J} > 1$ (50\%, 11 of 22) show inward or outward motions when compared with their subcritical counterparts (53\%, 18 of 34). We note, however, that a smaller fraction of protostellar cores with $M/M_J > 1$ (9 of 17, or 53\%) show contracting or expanding motions when compared with sub-critical protostellar cores showing contraction or expansion motions (13 of 13). In the protostellar case, core masses are continuously lowered as material accretes onto the central star, possibly reducing the core mass to Jeans mass ratio for more evolved (but still deeply embedded) protostellar cores. Perseus cores don't appear highly unstable and don't have large contraction motions, but previous studies show that some high mass regions also don't show large contraction motions despite appearing dynamically unstable \citep{Kauffmann2013}.

Most detected starless cores in our sample do not show direct evidence for inward nor outward motions (71\%, 20 of 28 cores). The fraction of starless cores having a measured contraction velocity is slightly larger for the supercritical case $M/M_\mathrm{J} > 1$ (2 of 5 cores, 40\%) compared to the subcritical case (5 of 21 cores, or 24\%). The overall fraction of supercritical starless cores showing contraction motions is thus similar to that found in Ophiuchus \citep[40\,\% vs.\ 40-60\,\%;][]{simpson11}, although Perseus does not harbor a large number of supercritical starless cores. No supercritical starless cores have measured outward motions. For the subcritical case, we find two starless cores with measured outward motions (B16, B34), along with three starless cores that have measured inward motion, a similar finding to \citet{Schnee2013}. In starless cores, outward motions have been attributed to oscillations of the core around an equilibrium \citep{lada03,lee11,broderick10} or to core expansion. Without spectral line maps of these two cores, however, we cannot distinguish between these scenarios. 

In summary, overall we find seven of 28 \HCO-detected starless cores that show evidence for contraction or expansion (25\,\% of the starless core sample) which more than doubles (15 of 28, or 54\,\%) for \HCO-detected protostellar cores. Thermally supercritical cores are somewhat more likely to show contraction motions compared to subcritical cores. Most starless cores in Perseus are likely not in a state of collapse (or expansion). \citet{Kirk2007} find that many cores in Perseus that overlap with our sample are consistent with being in virial equilibrium if the external pressure from the ambient cloud is taken into account. In their analysis, the authors model the cores as Bonnor-Ebert (BE) spheres to determine the external pressure. For a critical BE sphere, $M_\mathrm{BE} \sim 2 \ M_\mathrm{J}$. 

\subsection{Observed Core Contraction Speeds}
\label{subsec:speeds}

Where detected we find contraction speeds ranging between small subsonic values (0.03\,\kms) and large supersonic values (0.4\,\kms) at the typical temperatures of star-forming cores in Perseus \citep[$\sim 11$\,K;][]{Rosolowsky2008}. Of the six cores with contraction speeds greater than the sound speed at this typical temperature ($c_s =$ 0.2\,\kms), five are in clustered environments such as NGC 1333 and IC 348. In some cases, the large contraction speeds may be due to larger-scale collapse of regions like NGC 1333, but even in these clustered regions the contraction appears complex \citep{Walsh2006} and the contraction speeds are not uniform across the region. None of these sources are substantially super-Jeans, where one might expect large contraction speeds. In a couple cases where we find large contraction speeds, the uncertainties are also quite large. Such contraction sources are ideal candidates for further studies with greater sensitivity and in additional molecular tracers. We additionally find two sources whose contraction velocity is at the sound speed. Most of the contraction velocities are significantly less than the free-fall speed ($v_\mathrm{ff}$) for cores with the typical size and mass of the Perseus cores. We calculate the free-fall speeds as $v_\mathrm{ff} = - \sqrt{2GM/R_\mathrm{eff}}$. For contracting cores, $V_\mathrm{in} / v_\mathrm{ff}$ ranges from 0.04 to 0.97 and has an average value of 0.44, with one source (B104) having $V_\mathrm{in} / v_\mathrm{ff} = 1.22$. 


These contraction speeds are greater than predicted by ambipolar diffusion collapse models \citep{ciolek95}. In an analytic model of pressure-free collapse of centrally condensed cores, \citet{myers05} finds maximum contraction speeds of $\sim$ 0.1\,\kms~ at later times in the collapse ($t \sim 0.5\,t_\mathrm{ff}$, where $t_\mathrm{ff}$ is the free-fall time, and $t=0$ is the start of core collapse), and subsonic maximum contraction speeds over most of the collapse timescale. In the inside-out collapse model of \citet{Shu1977}, contraction velocities $\gtrsim 0.1$~\kms\, can be reached over the extent of a core, but only at later times when the protostar is visible at infrared wavelengths. Higher contraction speeds at outer core radii are predicted by hydrodynamic collapse models that converge to Larson-Penston flows \citep[e.g.,][]{foster93}; however, these highly supersonic contraction speeds are generally not observed in starless cores \citep{Keto2015}. 

In starless cores, contraction speeds are often subsonic when measured with molecular lines that trace similar densities to the {\HCO} transition observed here. For example, \citet{lee01} find typical contraction speeds {\Vin} $\sim$ 0.05 to 0.09\,\kms\ in emission from CS (2--1). Similarly to this study, \citet{Schnee2013} use {\HCOrot} to determine {\Vin} $\sim$ 0.1 to 0.2\,\kms. Notice that comparable contraction velocities have been measured at small scales with ALMA observations of IRAS 16293-2422 B \citep{Pineda2012, Zapata2013}. Both \citeauthor{lee01} and \citeauthor{Schnee2013} target well-studied starless cores within several nearby star-forming clouds (none of which overlap with our survey).

\edit{Analysis of hydrodynamic models of collapsing cores that are embedded in filaments has indicated that evidence for blue asymmetries in simulated rotational transitions of {\HCO} can be obscured by gas in the surrounding filament \citep{Chira2014}. \citeauthor{Chira2014} show that higher-order transitions are more accurate tracers of core contraction, because the line profiles are less confused by the emission of the surrounding lower density gas. They find that the fraction of cores with blue asymmetries observed in the {\HCOrot} transition is 67\,\%, with the remaining profiles being either red asymmetric or ambiguous because of confusion. Among the many regions within Perseus, this effect is most likely to be a factor in those like NGC 1333.}

Finally, we note that the sensitivity in this survey might not be high enough to detect large numbers of cores with very low contraction speeds. The ability of the {\Hill} model to determine the contraction speed of an asymmetric line profile accurately is strongly dependent on the S/N of the spectrum, and the model works best when the red and blue peaks are well separated, with an intensity difference between the red peak and self-absorption trough that is at least 10\,\% of the intensity of the blue-shifted peak. While we detect many of the cores in {\HCO} with S/N $> 10$, some sources with small contraction speeds, lower optical depths in the {\HCO}, or subtle line asymmetries would have large uncertainties in {\Vin} and so not be regarded as significant in our analysis. 

\subsection{Non-Thermal Motions Traced by Optically-Thin Line Widths}
\label{sec:widths}

While optically-thin lines do not show the asymmetric line profile indicative of contraction seen in optically-thick emission, any global contraction motion will result in observed line profiles that are wider than in cores that are not undergoing collapse or expansion \citep{myers05}, given similar thermal and non-thermal, non-contraction motions. Undetected contraction or expansion motions in classified ``static" cores might therefore result in a trend of increased line widths at larger {\MMJ} ratios. To explore this, we determined the non-thermal velocity dispersion, $\sigma_\mathrm{NT}$, for all \HCO-detected sources with optically-thin {\ammonia} line detections in \citet{Rosolowsky2008} using
\begin{equation}
\label{eq:dispersion}
{\sigma}_\mathrm{NT} = \sqrt{{\sigma}_\mathrm{thin}^2 - \frac{k_\mathrm{B} T_\mathrm{K}}{A m_\mathrm{H}} } \, .
\end{equation}
where $A$ is the mass number of the molecular species. As before, we set \tk\ based on \ammonia-derived gas temperatures and assume $T_\mathrm{K} = 11$~K for those sources without gas temperatures measured through {\ammonia} observations. We used {\ammonia} tracers instead of {\NtD} for the optically thin line width (${\sigma}_\mathrm{thin}$) to calculate the non-thermal velocity dispersion since the ammonia hyperfine structure is resolved which allows for more accurate measurements \citep{Rosolowsky2008}. The non-thermal velocity dispersion is listed for each core in Column 5 of Table~\ref{table:mm}. 

On average, cores with measured contraction/expansion motions have greater $\sigma_\mathrm{NT}$ than static cores, but there is no clear trend between $\sigma_\mathrm{NT}$ and $V_\mathrm{in}$. We show in Figure \ref{fig:mmcorrelations} (right) $\sigma_\mathrm{NT}$ as a function of \MMJ, separating the cores into those with detected contraction or expansion motions and static cores. Protostellar cores with measured contraction or expansion motions tend to have greater $\sigma_\mathrm{NT}$ with increasing \MMJ, but there are exceptions. No trend is seen, however, for starless or protostellar cores without measured contraction or expansion motions. Below the critical value $M/M_\mathrm{J} = 1$, both starless and protostellar cores have $\sigma_\mathrm{NT}$ values that range from subsonic to transonic (the sound speed for H$_2$ at 11~K is $c_s$ = 0.2\,\kms), although a larger fraction show subsonic $\sigma_\mathrm{NT}$, we still find many cores with supersonic $\sigma_\mathrm{NT}$. This is in contrast with \citet{Foster2009}, who find that almost all Perseus cores in their sample have subsonic $\sigma_\mathrm{NT}$. Above $M/M_\mathrm{J} = 1$, all static starless and protostellar cores, as well as starless collapsing cores, have subsonic $\sigma_\mathrm{NT}$, with one exception. 

Since we find no trend in $\sigma_\mathrm{NT}$ with {\MMJ} for the static cores, we conclude that any undetected contraction motions are small and subsonic. In the centrally-condensed, pressure-free collapse model of \citeauthor{myers05}, optically-thin line widths remain narrow at early times ($t \lesssim 0.3 t_\mathrm{ff}$), while collapse timescales are longer for the more realistic isothermal case by a factor $\sim 1.6$. This suggests that, as noted above, most of the starless cores in Perseus are either not currently in a state of collapse, or are in the very early stages where contraction speeds are low. 

\section{Summary}
\label{sec:summary}

We have performed a pointed molecular line survey of 91 dense starless and protostellar cores in Perseus. Observations of {\HCOrot} and {\NtDrot} rotational transitions were made using the James Clerk Maxwell Telescope, detecting 72 cores in {\HCO} and 24 in {\NtD}. 

\begin{enumerate}

\item We identify line asymmetries in optically-thick {\HCO} line profiles using a dimensionless asymmetry parameter, {\deltaV} \citep{Mardones1997}. We quantify line asymmetries for 20 cores \edit{that show no evidence of secondary cores along the line of sight (i.e., single cores)} and find that protostellar cores are more likely to exhibit line asymmetries than their starless counterparts.

\item We fit the \edit{line profiles of all \HCO-detected single cores} with an analytic collapse model \citep{DeVries2005} and determine the contraction velocity toward 22 cores that are well-fit by the model. We find only 7 of 28 (25\,\%) \edit{\HCO-detected single} starless cores that have spectra indicative of contracting or expanding motions and classify the remaining cores as ``static." Similarly, we find that 15 of 28 (54\,\%) \edit{\HCO-detected single} protostellar cores have spectra indicating inward or outward motions, with protostellar cores being far more likely than starless cores to show expanding motions, likely the result of expansions caused by embedded YSOs. Given a mean dust temperature $T_d = 11$\,K for Perseus, we find 6 sources that have contraction speeds greater than the sound speed of the molecular gas, significantly larger than previously observed contraction velocities toward starless cores. Of these sources 5 reside in clustered environments such as NGC 1333 and IC 348, suggesting that toward these cores the {\HCO} emission might be tracing global rather than local collapse motions. We additionally find 2 cores whose contraction speed is at the sound speed. Most of the cores have contraction velocities significantly less than their free-fall speeds. Only one source (B104) has a contraction speed of 1.22 times the free-fall velocity.

\item We investigate the relationship between the dimensionless asymmetry parameter ({\deltaV}) and the model-derived contraction (expansion) velocity ({\Vin}) and find that {\deltaV} is a good indicator of core contraction if the model-derived systemic velocity accurately traces the line profile center, \emph{when sources where multiple cores are likely overlapping along the line of sight are removed}. When using {\deltaV} as a tracer of core contraction, obtaining accurate {\Vthin} measurements is critical for identifying these overlapping sources, particularly in clustered regions. 

\item We determine the ratio of core mass to Jeans mass (\MJ). We find no overall trend between {\Vin} and \MMJ, but show that for both starless and protostellar sources, supercritical cores (where $M/M_\mathrm{J} > 1$) are somewhat more likely to have contraction or expansion motions.

\item In general, protostellar cores with known contraction or expansion motions also show increasing non-thermal line widths with \MMJ. This is not true for cores where we do not identify contraction or expansion-related line asymmetries. Furthermore, below $M/M_\mathrm{J} = 1$, static cores have both transonic and subsonic non-thermal line widths, while all static cores (with one exception) with $M/M_\mathrm{J} > 1$ have subsonic non-thermal line widths. This indicates that any undetected contraction or expansion motions are likely small and subsonic. Most starless cores in Perseus are therefore either not collapsing (or expanding), or are in the very early stages of collapse.

\end{enumerate}

The ability to study the complete kinematics of the Perseus cores is limited by the fact that these observations consist of single pointings. If the cores are evolving as described by the inside-out collapse model \citep{Shu1977}, then cores in the earliest stages of collapse might not experience contraction across the entire extent of the core and might go undetected if our pointing is off-center. Furthermore, it has been shown that the detectability of infall depends on the viewing angle \citep{Smith2012}. Future studies involving spectral line maps would be useful to test whether most Perseus cores are not in a state of collapse (or expansion), or if many are in the earliest stages of collapse that have gone undetected. Line maps would also help to disentangle the effects of local (core) and global collapse, as well as better identify molecular expansion in sources with negative contraction velocities.

\section*{Acknowledgments}

\edit{The authors thank the anonymous referee whose constructive suggestions have helped to improve the quality and clarity of this paper.} JLC was a summer research student at the Dunlap Institute for Astronomy and Astrophysics, University of Toronto, as part of the Summer Undergraduate Research Program. RKF is a Dunlap Fellow at the Dunlap Institute for Astronomy \& Astrophysics, funded through an endowment established by the Dunlap family and the University of Toronto. This work was supported in part by the Natural Sciences and Engineering Research Council of Canada. The James Clerk Maxwell Telescope has historically been operated by the Joint Astronomy center on behalf of the Science and Technology Facilities Council of the United Kingdom, the National Research Council of Canada, and the Netherlands Organisation for Scientific Research. The data were obtained under the proposal number M07BU33.

\begin{appendix}
\counterwithin{figure}{section} 
\section{}
\begin{figure*}
\includegraphics[width=18.5cm]{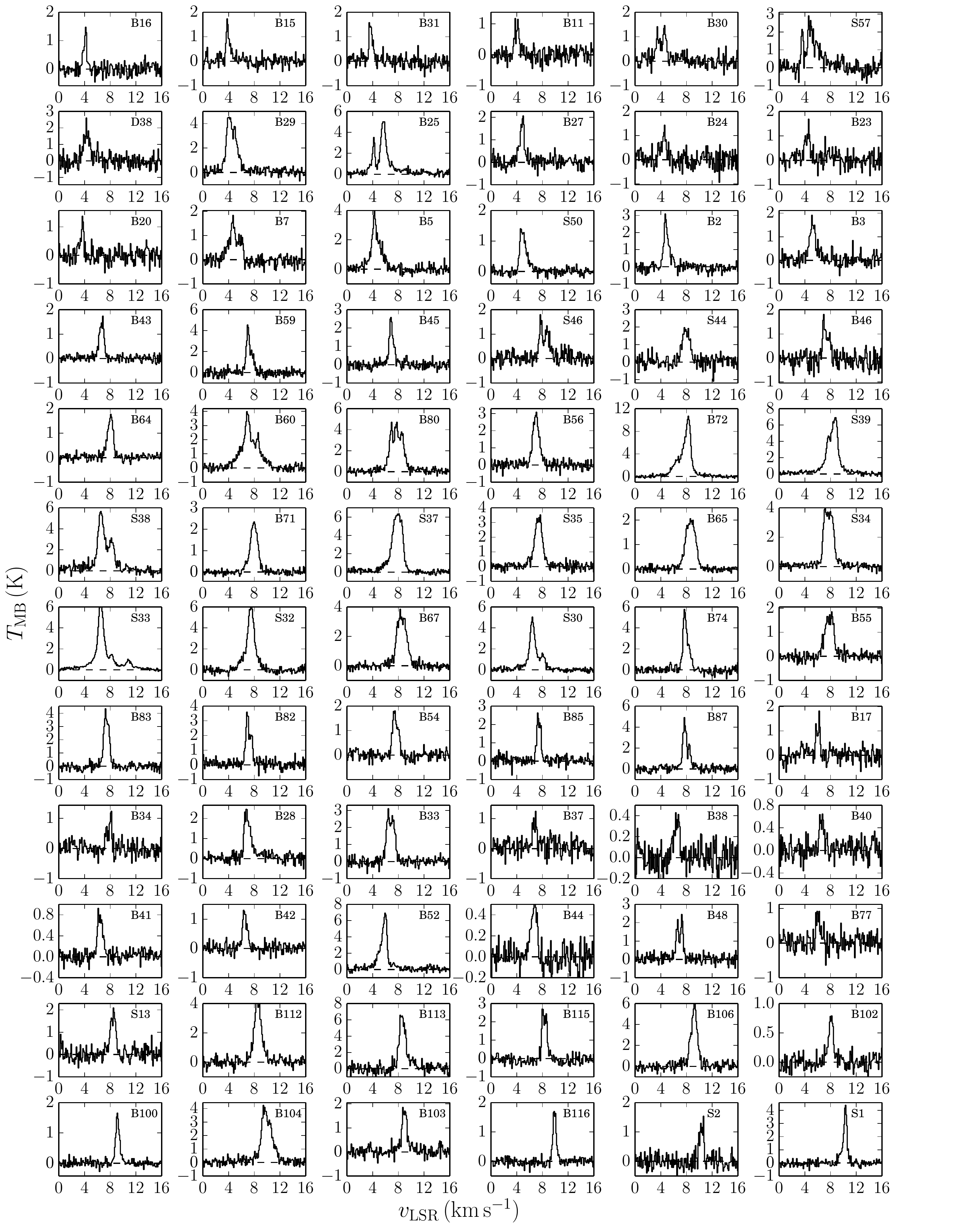}
\caption{Line profiles for 72 {\HCO} detections. The velocity scale is identical in all plots but the temperature scale differs.}
\label{fig:allhcoprofiles}
\end{figure*}

\begin{figure*}
\includegraphics[width=18.5cm]{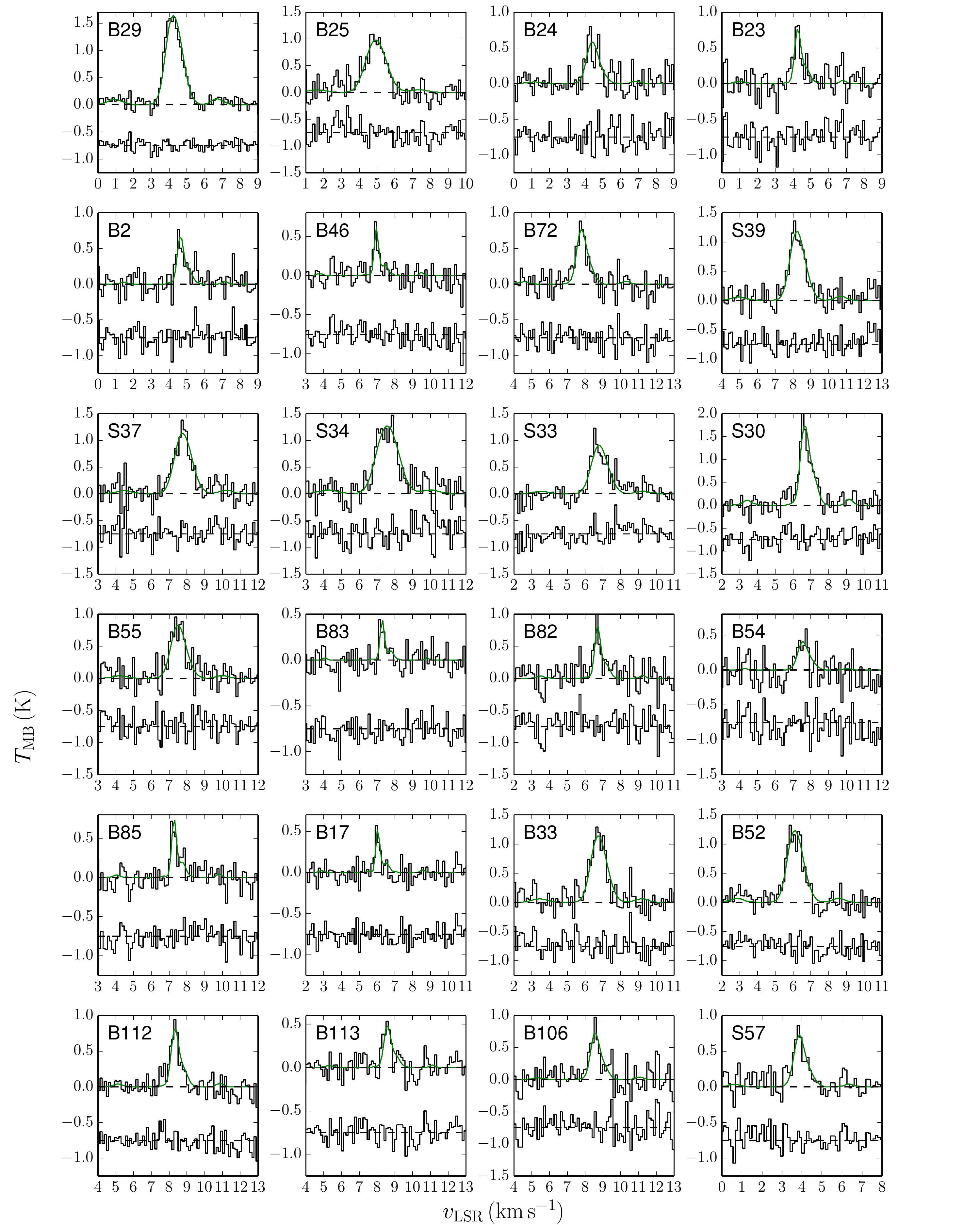}
\caption{Line profiles for 23 {\NtD} detections and one low S/N detection (B54, discussed in Section \ref{sec:results}) with the hyperfine structure best fits over-plotted (green). The resulting residuals are plotted below, displaced from centre on $-0.75$ K.}
\label{fig:alln2dprofiles}
\end{figure*}

Observed line profiles for all 72 detections of {\HCO} are shown in Figure \ref{fig:allhcoprofiles}, with sources in order of increasing RA as in Table~\ref{table:alltargetsinfo}.
Profiles for 23 detections of {\NtD} are shown in Figure \ref{fig:alln2dprofiles}, along with the profile for B54 where the {\NtD} line has S/N of only 2.3 but is well aligned with that of its {\HCO} counterpart. Hyperfine structure best fit results of {\NtD} are also shown.

\begin{figure*}
\includegraphics[width=20cm]{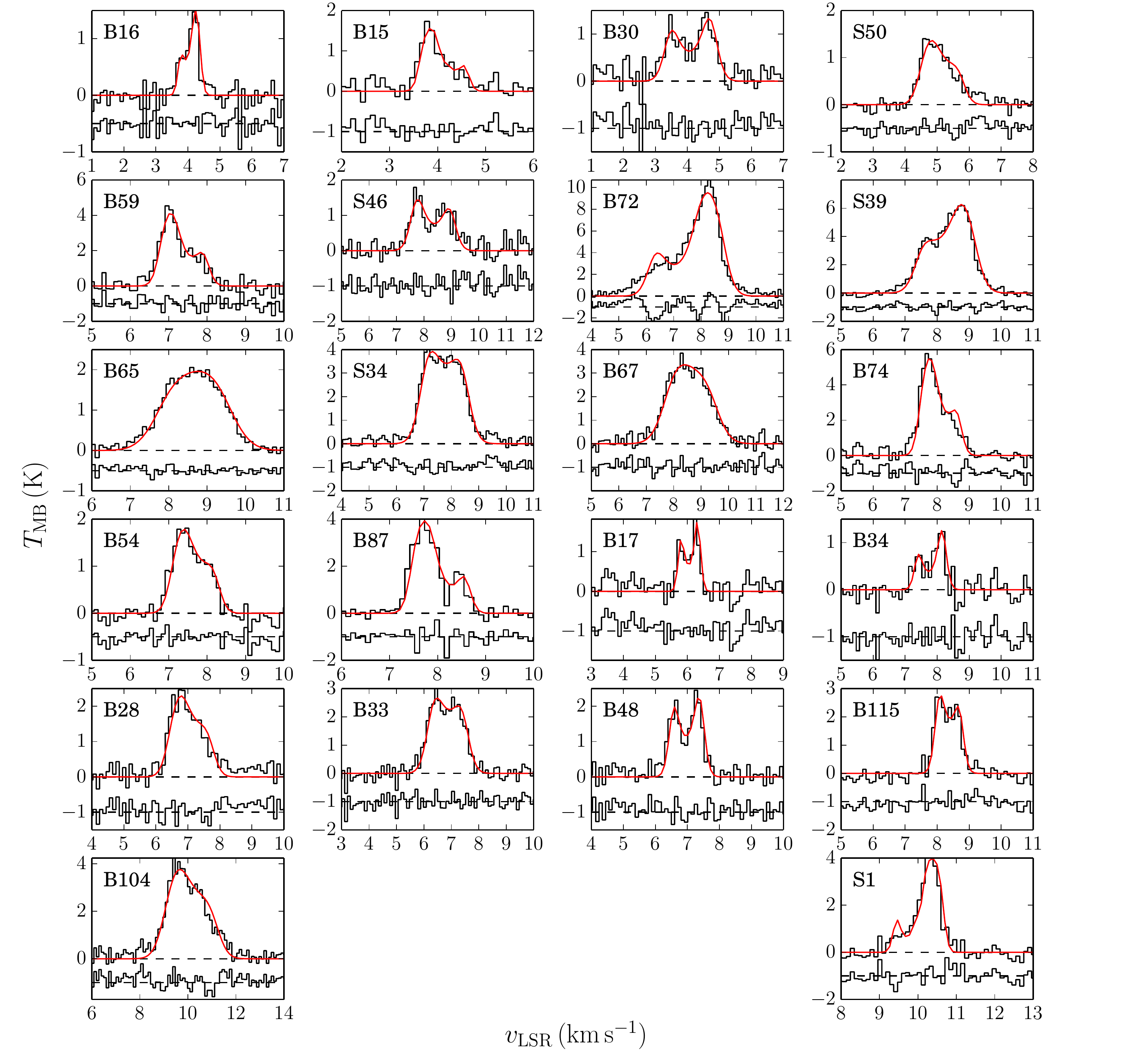}
\caption{Line profiles for 22 cores whose {\HCO} emission was modeled with the analytic {\Hill} model. The {\HCO} profile is shown along the 0 K baseline with the analytic model best fit over-plotted (red). The resulting residuals are plotted below, displaced from centre on $-1$ K.}
\label{fig:modelplots}
\end{figure*}

Figure \ref{fig:modelplots} shows the analytic fits to the 23 {\HCO} line profiles analysed with the {\Hill} model.

\end{appendix}
\newpage


\FloatBarrier
\bibliography{Biblio.bib}

\end{document}